%% file: ET_long.tex
\documentclass[%
 reprint,
 amsmath,amssymb,
 aps,
floatfix,
]{revtex4-2}

\usepackage{graphicx}
\usepackage{dcolumn}
\usepackage{bm}
\usepackage{natbib}
\usepackage{subfig}
\usepackage{xcolor}
\usepackage{hyperref}

\begin{document}

\preprint{APS/123-QED}

\title{Constraining parameters of low mass merging compact binary systems with Einstein Telescope alone}

\author{Neha Singh}
\email{singh@lapth.cnrs.fr}
\affiliation{Laboratoire d’Annecy-le-Vieux de Physique Théorique (LAPTh), USMB, CNRS,\\ F-74940 Annecy, France \\ 
Astronomical Observatory, University of Warsaw,\\ Aleje Ujazdowskie 4, 00-478 Warsaw, Poland}
\author{Tomasz Bulik}
\email{tb@astrouw.edu.pl}
\affiliation{Astronomical Observatory, University of Warsaw,\\ Aleje Ujazdowskie 4, 00-478 Warsaw, Poland}

\date{\today}
\begin{abstract}
The Einstein Telescope (ET), a future third-generation  gravitational wave detector will have detection sensitivity for gravitational wave signals down to 1 Hz. This improved low-frequency sensitivity of the ET will allow the observation of low mass binaries for a longer period of time in the detection band before their merger. Because of an improved sensitivity as compared to current and advanced 2G detectors, the detection rate will also be greatly improved. Given the high detection rate of merging compact binaries with the ET, it will be a useful instrument to conduct population studies. In this paper we present an algorithm to estimate the parameters of the low mass merging compact binary systems such as localization, chirp mass, redshift, mass ratios and total mass of the source which are crucial in order to estimate the capability of the ET to study various compact binary populations. For the compact binary population distributed uniformly in comoving volume we find that with single ET, $\approx 1\%$ of binaries can be localized within 800 square degrees. The values of chirp mass and total mass can be constrained within $\lesssim 5\%$ error, while $z$ and $D_L$ can be estimated with an error of $\lesssim 15\%$ for effective SNR $\gtrsim 50$ using single ET.

\end{abstract}


\maketitle

\input{macros}

\section{Introduction}

The second-generation gravitational wave (GW) detectors initiated GW astronomy with the first detection of merging compact objects GW150914 \cite{2016PhRvL.116f1102A}, which was the direct detection of a binary black hole (BBH). Numerous BBH detections in the next run since then have shown the existence of a population of stellar-mass black holes (BHs) undetected in previous observations which is much heavier than those detected through the observation of x-ray binaries \cite{2021PhRvX..11b1053A,2019PhRvX...9c1040A,2020arXiv200408342T,2020PhRvL.125j1102A}. The first detection of a binary neutron star (BNS) inspiral \cite{2017PhRvL.119p1101A} with simultaneous gamma-ray burst observation and the subsequent detection of the electromagnetic counterpart provided a better understanding of the origin of short gamma-ray bursts \cite{2017ApJ...848L..14G,2017ApJ...848L..15S,2017ApJ...848L..13A,2017Sci...358.1556C,2017Natur.551...71T,2017Sci...358.1579H,2017ApJ...848L..12A}. The observations of the associated kilonova proved that the BNS mergers source the formation of the heaviest metals through r-process nucleosynthesis. The joint detection of GWs and gamma-ray bursts proved the speed of GW and the speed of light to be equal with an accuracy of 1 in $10^{15}$ \cite{2017ApJ...848L..13A}. 

Detailed studies of GW sources in the Universe  will be continued with the 3G detectors such as ET \cite{2011CQGra..28i4013H,2010CQGra..27s4002P} or Cosmic Explorer (CE) \cite{PhysRevD.91.082001,Abbott2017,2019BAAS...51g..35R}. ET is planned to have a  detection sensitivity down to 1 Hz \cite{2008arXiv0810.0604H,2012CQGra..29l4006H}. This is required so as to have the ability to detect BBHs with components of higher mass such as $10^2–10^4 M_{\odot}$ \cite{2011PhRvD..83d4020H,2011PhRvD..83d4021H,2011GReGr..43..485G,2010ApJ...722.1197A} which are yet to be detected.

Given the improvement in sensitivity by a factor of 10 in the intermediate frequency range, and several orders of magnitude improvement in the low-frequency band, as compared to 2G detectors, ET will be able to detect systems such as  BBH  and neutron star - black hole (NSBH) binaries of total mass in the range of $20 - 100\; M_{\odot}$, up to redshift $z\sim 20$; high mass BHs with masses of the order $10^3 \; M_{\odot}$ up to $z \sim (1-5)$; BNS binaries of total masses $\sim 3 \; M_{\odot}$ up to $z \sim (2-3)$  \cite{Sathyaprakash:2019nnu, 2020JCAP...03..050M}. The detection rate will also be greatly improved as compared to the current and advanced 2G detectors. The expected detection rates based on the ET-D \cite{2011CQGra..28i4013H} design sensitivity are  $\sim 10^5 - 10^6$ BBH detections and $\sim 7 \times 10^4$ BNS detections in one year \cite{2012PhRvD..86l2001R,2014PhRvD..89h4046R, 2019JCAP...08..015B}.

Because of an improved low-frequency sensitivity, ET will observe low mass binaries for a longer period of time in the detection band before their merger. With the lowest frequency detection sensitivity for ET-D being down to 1Hz, the BNS signals can stay in the detectable band from a few minutes to several days. Thus, it is necessary to take into account the effect of rotation of Earth on the response function. \citet{2018PhRvD..97f4031Z} and \citet{2018PhRvD..97l3014C} have done such an analysis using a Fisher information matrix to study the effects of the time-dependent detector response due to Earth’s rotation on long-duration signals from systems such as BNS and NSBH for estimating the uncertainties in the measurement of signal parameters using 3G detectors such as ET and CE. In the previous work \cite{2021PhRvD.104d3014S}, we presented an algorithm to localize and constrain the parameters of BBH coalescences such as the chirp mass $\mathcal{M}$ and redshift $z$ or luminosity distance $D_L$ using single ET. In this work, we continue to explore the capability of ET as a single instrument to study longer-duration signals from coalescing low mass compact binary systems and present a simplified approach to estimate the parameters of low mass merging compact binary systems. Given the high detection rate of merging compact binaries with ET, it will be a useful instrument for conducting population studies. We present an algorithm to estimate the angles describing the location of the source, the inclination, and polarization of an inspiralling compact binary system using the ratios of the signal to noise ratios (SNRs) generated in each of the three detectors in single ET. We also demonstrate that single ET can break the chirp mass - redshift degeneracy and, thus, provide estimates of the chirp mass, redshift, mass ratios, and total mass of the source. These estimates are crucial in order to estimate the capability of ET to study various compact binary populations \cite{2004A&A...415..407B}. 

\section{ET as a single instrument}\label{sec:ET_single}

The ET will comprise of three coplanar detectors of equal arm length of 10 km, aligned in the form of an equilateral triangle so that the opening angle will be $60 ^{\circ}$ and it will use Michelson interferometry. Multiple design configurations have been studied over time. The first basic design considered was ET-B \cite{2008arXiv0810.0604H}. It was based on a single cryogenic interferometer and covered the full frequency range of interest. It was then updated to a xylophone design resulting in the ET-C sensitivity \cite{2012CQGra..29l4006H} in which each detector consisted of two interferometers, each with an opening angle of $60^{\circ}$, with one optimized for low frequencies and the other optimizsed for high frequencies. ET-D \cite{2011CQGra..28i4013H} is a realistic version of ET-C since it considers an improved noise model. Each ET detector will have two interferometers, one each for the low and high frequencies. The final triangular design of the ET will have three such detectors and so six interferometers in total.

While observing a GW signal which stays in the detection band of the detector for a long duration one has to take into account the change in the antenna response with the rotation of Earth. Following the detailed treatment given in \citet{PhysRevD.58.063001}, which takes into account the motion of Earth, the time-dependent antenna response function for a single detector in the reference frame of the celestial sphere at time $t$ is given as

\begin{subequations}\label{antenna_long}
\begin{equation}
    F_{+}(t) = \sin \eta \left[ a(t) \cos 2 \psi + b(t) \sin 2 \psi \right]
\end{equation}

\begin{equation}
    F_{\times}(t) = \sin \eta \left[ b(t) \cos 2 \psi - a(t) \sin 2 \psi \right]
\end{equation}
\end{subequations}
where,

\begin{subequations}
\begin{equation}
\begin{split}
    a(t)& = \frac{1}{16} \sin 2 \gamma (3 - \cos 2 \lambda)(3 - \cos 2 \delta)\cos[2(\alpha - \phi_r - \Omega_rt)]\\
    & - \frac{1}{4} \cos 2 \gamma \sin \lambda (3 - \cos 2 \delta) \sin[2(\alpha - \phi_r - \Omega_rt)]\\
    & + \frac{1}{4}\sin 2\gamma \sin 2 \lambda \sin 2 \delta \cos[\alpha - \phi_r - \Omega_rt]\\
    & - \frac{1}{2}\cos 2\gamma \cos \lambda \sin 2 \delta \sin[\alpha - \phi_r - \Omega_rt]\\
    & +\frac{3}{4} \sin 2 \gamma \cos^2 \lambda \cos^2 \delta
\end{split}
\end{equation}
and

\begin{equation}
\begin{split}
    b(t) & = \cos 2 \gamma \sin \lambda \sin \delta \cos[2(\alpha - \phi_r - \Omega_rt)]\\
    & + \frac{1}{4} \sin 2 \gamma (3 - \cos 2 \lambda)\sin \delta \sin[2(\alpha - \phi_r - \Omega_rt)]\\
    & + \cos 2 \gamma \cos \lambda \cos \delta \cos[\alpha - \phi_r - \Omega_rt]\\
    & + \frac{1}{2} \sin 2 \gamma \sin 2 \lambda \cos \delta \sin[\alpha - \phi_r - \Omega_rt]
\end{split}
\end{equation}
\end{subequations}
where $\alpha$ is the right  ascension, $\delta$ is the declination of the GW source, $\psi$ is the polarization angle, and $\lambda$ is the latitude for the detector location. $\Omega_r$ is Earth's rotational angular velocity and $\phi_r$ is the phase defining the position of Earth in its diurnal motion at $t = 0$. The quantity $(\phi_r + \Omega_rt)$ is the local sidereal time at the detector site, measured in radians, while $\gamma$ determines the orientation of the detector arms and is measured counter-clockwise from East to the bisector of the interferometer arms. Finally, $\eta$ is the angle between the interferometer arms. In the case of ET,  $\eta = 60^{\circ}$.

Using Eq. (\ref{antenna_long}), the antenna response functions can be calculated for any given instant of time $t$. We use the currently planned  design of ET-D \cite{2011CQGra..28i4013H}, consisting of three overlapping detectors, arranged in an equilateral configuration with arm-opening angles of $60^{\circ}$. The location of ET detector for our analysis is chosen to be at the Virgo site \cite{2012JInst...7.3012A,2015CQGra..32b4001A}.

\section{Signal characteristics of coalescing compact binary systems}\label{sec:signal_long}

We consider ET as a single instrument rather than a part of network to detect the gravitational radiation from an inspiralling compact binary system in this work. Inspiralling compact binary systems are also known as chirping binaries. The two polarizations of the GW signal from such a system have a monotonically increasing frequency and amplitude with the orbital motion radiating away GW energy. If $t_c$ is the time of the termination of the waveform, then the two polarizations $h_{+}$ and $h_{\times}$, of the waveform at time $t< t_c$ for a binary with the chirp mass $\mathcal{M}$, merging at a distance $D_{L}$ (described in Sec 3 in Ref. \cite{Findchirp}) are given as

\begin{subequations}\label{hpluscross_ant}

\begin{equation}
\begin{split}
    h_{+}(t) = -\frac{1+\cos^{2}\iota}{2}\left(\frac{G \mathcal{M} }{c^2D_{L}}\right)\left(\frac{t_c-t}{5G\mathcal{M} /c^3}\right)^{-1/4}\\
    \times \cos\left[2\Phi_c + 2\Phi \left(t-t_c ; M,\mu\right)\right],
\end{split}
\end{equation}
\begin{equation}
\begin{split}
    h_{\times}(t) = -\cos\iota \left(\frac{G \mathcal{M}}{c^2D_{L}}\right)\left(\frac{t_c-t}{5G\mathcal{M}/c^3}\right)^{-1/4}\\
    \times \sin\left[2\Phi_c + 2\Phi \left(t-t_c ; M,\mu\right)\right],
\end{split}
\end{equation}
\end{subequations}
where $c$ is the speed of light, $G$ is the gravitational constant and $\iota$ is the angle of inclination of the orbital plane of the binary system with respect to the observer. $\mu$ is the reduced mass of the binary system. The angle $\Phi \left(t-t_c ; M,\mu\right)$ gives the orbital phase of the binary system. The chirp mass $\mathcal{M}$ for a binary system composed of component masses $m_1$ and $m_2$ is defined as $\mathcal{M} = (m_1m_2)^{3/5}/M^{1/5}$, where $M = m_1+m_2$, is the total mass and $\Phi_{c}$ is the phase of the termination of the waveform \cite{Findchirp}. The strain in the detector is given as

\begin{equation}\label{h_t_ant}
    h(t)= F_+ (t)h_+(t+t_c-t_0) + F_{\times}(t) h_\times(t+t_c-t_0)
\end{equation}
where $F_+$ and $F_\times$ are the antenna response function of one of the three detectors in ET, as defined in Eq. (\ref{antenna_long}),  $t_0$ is the time of coalescence in the detector frame and $(t_0 - t_c)$ is the travel of time from the source to the detector. Substituting the values of the two polarizations from Eq. (\ref{hpluscross_ant}) in Eq. (\ref{h_t_ant}) gives the value of strain

\begin{equation}\label{ht}
\begin{split}
    h(t)= -\left(\frac{G\mathcal{M}}{c^2}\right)\left(\frac{\Theta}{4D_L}\right)\left(\frac{t_0-t}{5G\mathcal{M}/c^3}\right)^{-1/4}\\
    \times \cos\left[2\Phi_0 + 2\Phi \left(t-t_c ; M,\mu\right)\right]
\end{split}
\end{equation}
where
\begin{equation}\label{theta}
    \Theta\equiv 2 \left[F_{+}^{2}\left(1+\cos^{2}\iota\right)^{2} + 4F_{\times}^{2}\cos^{2}\iota\right]^{1/2}
\end{equation}
and

\begin{equation}\label{snr_phase}
    2\Phi_0 = 2\Phi_c - \arctan\left(\frac{2F_\times\cos\iota}{F_+\left(1 + \cos^2\iota\right)}\right)
\end{equation}
with $0<\Theta<4$.

In a given duration, for which it can be assumed that the time of the signal in the detector bandwidth is short enough to ignore the change in the antenna response functions of the detector due to rotation of Earth, the Fourier transform of the GW signal amplitude $h(t)$ in terms of frequency $f$ is \cite{PhysRevD.44.3819,TaylorGair2012,2010ApJ...716..615O}

\begin{equation}
    |\tilde{h}(f)|= \frac{2c}{D_L}\left(\frac{5G\mu}{96c^3}\right)^{1/2}\left(\frac{GM}{\pi^2c^3}\right)^{1/3}\left(\frac{\Theta}{4}\right)f^{-7/6}\label{h_fourier}.
\end{equation}
The SNR $\rho_j$ for $j = (1,2,3)$ for each of the three ET detectors, obtained using match-filtering assuming that they have identical noise is given as \cite{TaylorGair2012,LeeFinn96}

\begin{equation}\label{snr}
\rho_j \approx 8 \Theta_j \frac{r_{0}}{D_{L}}
\left(\frac{\mathcal{M}_{z}}{\mathcal{M}_{BNS}}\right)^{5/6}\sqrt{\zeta\left(f_{max}\right)} 
\end{equation}
where  $\mathcal{M}_{z}= (1+z)\mathcal{M} $ is the redshifted chirp mass.  $\mathcal{M}_{BNS}\approx 1.218 M_\odot $  is the chirp mass of an equal mass binary with each component mass being $1.4 M_{\odot}$, while

\begin{equation}\label{zetafunc}
\zeta\left(f_{max}\right) = \frac{1}{x_{7/3}}\int^{2f_{max}}_{1}\frac{df \left( \pi M_{\odot}\right)^{2}}{\left(\pi f M_{\odot}\right)^{7/3}S_{h}\left(f\right)}
\end{equation}
where $S_{h}\left(f\right)$ is the power spectral density (PSD) for ET-D configuration for the ET-D noise curve \cite{2011CQGra..28i4013H} and

\begin{equation}\label{x_7_3}
x_{7/3} = \int^{\infty}_{1\rm Hz}\frac{df \left( \pi M_{\odot}\right)^{2}}{\left(\pi f M_{\odot}\right)^{7/3}S_{h}\left(f\right)}.
\end{equation}
The characteristic distance sensitivity $r_0$ is

\begin{equation}\label{detreach}
r^{2}_{0} = \frac{5}{192 \pi}\left(\frac{3 G}{20}\right)^{5/3}x_{7/3}\frac{M^{2}_{\odot}}{c^{3}},
\end{equation}
The frequency at the end of the inspiral phase $f_{max}$, is given as

\begin{equation}\label{fmax}
f_{max} = 785\left(\frac{M_{BNS}}{M(1+z)}\right)  \;\rm Hz
\end{equation}
where $M_{BNS}=2.8 M_\odot$ is the total mass of an equal mass binary with each component mass of $1.4 M_{\odot}$. We can define the combined effective SNR for the combined signal from three detectors as:

\begin{equation}\label{snreff}
\rho_{eff} = 8 \Theta_{eff} \frac{r_{0}}{D_{L}}\left(\frac{\mathcal{M}_{z}}{1.2 M_{\odot}}\right)^{5/6}\sqrt{\zeta\left(f_{max}\right)} 
\end{equation}
where the effective antenna response function  $\Theta_{eff}$ is

\begin{equation}\label{thetaeff}
\Theta_{eff} = \left(\Theta_{1}^{2} + \Theta_{2}^{2} + \Theta_{3}^{2}\right)^{1/2}.
\end{equation}

\section{The plan of the analysis}\label{sec:plan_long}

We use the response functions given by Eq. (\ref{antenna_long}), as defined in Sec. \ref{sec:ET_single}, in the Celestial sphere frame of reference \cite{PhysRevD.58.063001}, for this long-duration signal analysis. Assuming that the response functions do not change much during 5 min, we divide the inspiral signal into 5 minutes segments from the time it enters the detection band at 1 Hz. The duration of the last segment  will be $\leq$ 5 min, since it is limited by $f_{max}$, the frequency at the end of the inspiral, given by Eq. (\ref{fmax}). 

We assume that in the case of detection of a coalescing low mass binary system, the observables $D$ are as follows: (a) The three SNRs $\rho^i_j$ defined by Eq. (\ref{snr}) for each $i^{th}$ segment of the signal. (b) The phase of the strain $\Phi^i_{o,j}$ for $j = (1,2,3)$ corresponding to three detectors, defined in Eq. (\ref{snr_phase}) for each $i^{th}$ segment of the signal. The quantity $\Phi_0$ is the best match phase obtained by maximizing the matched-filter output over the phase of the strain $h(t)$. The details are given in Ref. \cite{Findchirp}. (c) The GW frequency at the start and end of each segment of the detected signal. (d) The redshifted chirp mass $\mathcal{M}_{z}$. (e) The frequency at the end of the inspiral, corresponding to the innermost stable circular orbit, $f_{max}$. 

The observed GW frequency  $f^{obs}_{gw}$ can be calculated using Eq. (4.195) in \citet{Maggiore:1900zz}

\begin{equation}\label{f_gw}
f^{obs}_{gw} = \frac{1}{\pi} \left( \frac{5}{256}\frac{1}{\tau_{obs}}  \right)^{3/8} \left( \frac{G\mathcal{M}_{z}}{c^3}  \right)^{-5/8}
\end{equation}
where $\tau_{obs}$ is the time to coalescence measured in the observer's frame. The minimum frequency for the detection sensitivity of the detector and the frequency $f_{max}$ sets the limit on $\tau_{obs}$ spent in the detection band.

If $\tau_{i-1}$ and $\tau_{i}$ are the start and end values of $\tau_{obs}$ respectively, for the $i^{th}$ segment then the corresponding values $f_{i-1}, f_{i}$ of $f^{obs}_{gw}$ will be

\begin{equation}
f_{i-1} = \frac{1}{\pi} \left( \frac{5}{256}\frac{1}{\tau_{i-1}}  \right)^{3/8} \left( \frac{G\mathcal{M}_{z}}{c^3}  \right)^{-5/8}
\end{equation}
and
\begin{equation}
f_{i} = \frac{1}{\pi} \left( \frac{5}{256}\frac{1}{\tau_{i}}  \right)^{3/8} \left( \frac{G\mathcal{M}_{z}}{c^3}  \right)^{-5/8}.
\end{equation}
In order to constrain the angles defining the strain in the detector, we use the ratios of SNR in each segment. The SNR for $i^{th}$ segment in $j^{th}$ detector can be written using Eq. (\ref{snr}) as:

\begin{equation}\label{snr_seg}
\rho^i_j \approx 8 \Theta^i_j \frac{r_{0}}{D_{L}}
\left(\frac{\mathcal{M}_{z}}{\mathcal{M}_{BNS}}\right)^{5/6}\sqrt{\zeta^i\left(f_{i-1}, f_{i}\right) }
\end{equation}
where,

\begin{equation}\label{zetafunc_seg}
\zeta^i\left(f_{i-1}, f_{i}\right) = \frac{1}{x_{7/3}}\int^{f_i}_{f_{i-1}}\frac{df \left( \pi M_{\odot}\right)^{2}}{\left(\pi f M_{\odot}\right)^{7/3}S_{h}\left(f\right)}.
\end{equation}
and 

\begin{equation}\label{theta_seg}
    \Theta^i_j\equiv 2 \left[(F^i_{+})^{2}\left(1+\cos^{2}\iota\right)^{2} + 4(F^i_{\times})^{2}\cos^{2}\iota\right]^{1/2}_j
\end{equation}
where $F^i_{+}$ and $F^i_{\times}$ are the antenna response functions for the $j^{th}$ detector in $i^{th}$ segment. The effective SNR for the $i^{th}$ segment is

\begin{equation}\label{snreff_seg}
\rho^i_{eff} = 8 \Theta^i_{eff} \frac{r_{0}}{D_{L}}\left(\frac{\mathcal{M}_{z}}{1.2 M_{\odot}}\right)^{5/6}\sqrt{\zeta^i\left(f_{i-1}, f_{i}\right)} 
\end{equation}
where

\begin{equation}
(\rho^i_{eff})^2 = (\rho^i_1)^2 + (\rho^i_2)^2 + (\rho^i_3)^2 
\end{equation}
and the function  $\Theta^i_{eff}$ is
\begin{equation}\label{thetaeff_seg}
(\Theta^i_{eff})^2 = (\Theta^i_{1})^{2} + (\Theta^i_{2})^{2} + (\Theta^i_{3})^{2}.
\end{equation}
Using Eq. (\ref{snr_phase}), we can write for $i^{th}$ segment in $j^{th}$ detector

\begin{equation}\label{snr_phase_seg}
    2\Phi^i_{0,j} = 2\Phi_c - \arctan\left(\frac{2F^i_{\times j}\cos\iota}{F^i_{+j}\left(1+\cos^2\iota\right)}\right)
\end{equation}

The antenna response function for the detector is dependent on ($\delta, \alpha, \psi$) as seen from Eq. (\ref{antenna_long}) where $\delta$ is the declination, $\alpha$ is the right ascension for the location of the binary on the celestial sphere, and $\psi$ is the polarization angle. Thus the quantity $\Theta$ depends on the four angles ($\delta, \alpha, \iota, \psi$), as seen from Eq. (\ref{theta}) where $\iota$ is the inclination angle of the binary with respect to the direction of observation. We choose $\cos\delta, \alpha/ \pi$, $\cos \iota$, and $\psi/ \pi$ to be uncorrelated and distributed uniformly over the range $[-1,1]$.

We use the SNRs in each segment for each of the three detectors in the triangular configuration of ET, to constrain the value of the effective antenna pattern.
The value of the ratios of the SNRs are denoted as:  $\rho^i_{21} \equiv \rho_2/\rho_1$, and $\rho^i_{31} \equiv \rho_3/\rho_1$ in the $i^{th}$ segment. We note that using Eq. (\ref{snr_seg}) for each segment, these ratios are

\begin{equation}\label{theta-ratios}
     \rho^i_{21} = \frac{\Theta^i_2}{\Theta^i_1}\equiv\Theta^i_{21} \ \ \ {\rm and}\ \ \ \ \   \rho^i_{31} = \frac{\Theta^i_3}{\Theta^i_1}\equiv \Theta^i_{31},
\end{equation}
The difference of the phase $\Phi^i_0$ using Eq. (\ref{snr_phase}) for the three detectors in the $i^{th}$ segment is

\begin{equation}\label{phi-diff}
    \Phi^i_{21} = \Phi^i_{0,2} - \Phi^i_{0,1} \ \ \ {\rm and}\ \ \ \ \   \Phi^i_{31} = \Phi^i_{0,3} - \Phi^i_{0,1}
\end{equation}
The quantities $\rho^i_{21}, \rho^i_{31}, \Phi^i_{21},$ and $\Phi^i_{31}$ depend only on the position on the sky, polarization and inclination angle of the binary system. Thus we have constraints on these four angles from Eqs. (\ref{theta-ratios}) and (\ref{phi-diff}) for each segment. In this analysis we assume that the measurement errors on the SNRs is Gaussian with the standard deviations for $\rho^i_j$ and $\Phi^i_{j}$ being $\sigma_{\rho}=1$, and $\sigma_{\Phi}=\pi/\rho$, respectively. Note that this is a conservative assumption in comparison to the errors on the SNRs of the recent GW detections, mentioned in GWTC-2 \cite{2021PhRvX..11b1053A, 2021arXiv211103606T}.

The probability densities of the measured SNR ratios $P(\rho^i_{kj})$ and difference between the orbital phase $P(\Phi^i_{kj})$ between the $k^{th}$ and $j^{th}$ detector for the $i^{th}$ segment are

\begin{equation}
P(\rho^i_{kj}) = \int d\rho^i_k d\rho^i_j \delta(\rho^i_{kj}-\rho^i_k/\rho^i_j) P(\rho^i_k) P(\rho^i_j),
\end{equation}
and

\begin{equation}
P(\Phi^i_{kj})  = \int d\Phi^i_k d\Phi^i_j \delta(\Phi^i_{kj}-(\Phi^i_k-\Phi^i_j)) P(\Phi^i_k) P(\Phi^i_j).
\end{equation}

We start the analysis of the inspiral signal by constraining the angles in each of the $i^{th}$ segment of the signal. In order to do so, we consider the first set of data $D_1$ for the $i^{th}$ segment to be $D_{1i} \equiv (\rho^i_{21}, \rho^i_{31}, \Phi^i_{21} \Phi^i_{31})$. We use Bayes theorem in each segment to obtain the constraints on the position in the sky and on the polarization and inclination angles. For $\Omega_{sky}\equiv (\delta,\alpha)$, $\Omega_{source} \equiv (\iota, \psi)$, we have

\begin{equation}\label{fourangles_seg}
    P(\Omega_{eff}|D_{1i}, I) = \frac{P(\Omega_{eff}|I) P(D_{1i}| \Omega_{eff},I)}{P(D_{1i}|I)}
\end{equation}
where $\Omega_{eff} \equiv (\Omega_{sky}, \Omega_{source})$. 
Since the prior probability $P(\Omega_{eff}|I)$ is uniform on both the source and the detector sphere we can write the prior probability as

\begin{equation}\label{omega_prior_seg}
    P(\Omega_{eff}|I) = \frac{1}{(4\pi)^2}
\end{equation}
and the likelihood as

\begin{equation}
\begin{split}
    P( D_{1i} | \Omega_{eff},I)  = \int d\rho^i_{21} P(\rho^i_{21}) 
\int d\rho^i_{31}P(\rho^i_{31})\\
 \times \int d\Phi^i_{21} P(\Phi^i_{21}) \int d\Phi^i_{31}P(\Phi^i_{31}) \\
 \times \delta\left(\rho^i_{21}-\rho^i_{21}(\Omega_{eff})\right) \delta\left(\rho^i_{31}-\rho^i_{31}(\Omega_{eff})\right)\\
 \times \delta\left(\Phi^i_{21}-\Phi^i_{21}(\Omega_{eff})\right) \delta\left(\Phi^i_{31}-\Phi^i_{31}(\Omega_{eff})\right)
\end{split}
\end{equation}
Given the information on the sky localization and source angles in the $i^{th}$ segment we constrain $\Theta^i_{eff}$ by marginalizing over $\Omega_{eff}$ as

\begin{equation}\label{thetaprob_seg}
\begin{split}
P(\Theta^i_{eff}|& D_{1i}, I)  = \int d\Omega_{eff} P(\Theta^i_{eff}, \Omega_{eff}|D_{1i}, I)\\
& = \int d\Omega_{eff} P(\Omega_{eff}|D_{1i}, I)P(\Theta^i_{eff}|\Omega_{eff},D_{1i}, I)\\
& = \int d\Omega_{eff} P(\Omega_{eff}|D_{1i}, I)P(\Theta^i_{eff})\\
& \; \; \; \times \delta\left(\Theta^i_{eff} - \Theta^i_{eff}(\Omega_{eff})\right)
\end{split}
\end{equation}
We assume that,

\begin{equation}\label{thetaeffprior_seg}
    P(\Theta^i_{eff}) = \frac{1}{\Delta \Theta_{eff}^{max}}
\end{equation} 
where $\Theta_{eff}^{max} = 6$. This is so because, for each of the three detectors in the ET, the angle between the interferometer arms is $60^{\circ}$. We can thus rewrite Eq. (\ref{thetaprob_seg}) as

\begin{equation}\label{thetaeffprob_seg}
\begin{split}
    P(\Theta^i_{eff}|D_{1i}, I) = \frac{1}{\Delta \Theta_{eff}^{max}} \int & d\Omega_{eff} P(\Omega_{eff}|D_{1i}, I) \\
    & \times \delta\left(\Theta^i_{eff} - \Theta^i_{eff}(\Omega_{eff})\right)
\end{split}
\end{equation}
Substituting Eq. (\ref{fourangles_seg}) in Eq. (\ref{thetaeffprob_seg}) gives the probability density for $\Theta^i_{eff}$. Moving further, by rearranging Eq. (\ref{snreff_seg}), we notice that

\begin{equation}\label{lambda1}
    \frac{\Theta^i_{eff} \sqrt{\zeta^i\left(f_{i-1}, f_{i}\right)}}{\rho^i_{eff}}   \approx \left(\frac{8r_{0}}{D_{L}}\left(\frac{\mathcal{M}_{z}}{\mathcal{M}_{BNS}}\right)^{5/6} \right)^{-1}.
\end{equation}
It is seen that, while the quantities $\Theta^i_{eff}$, $\zeta^i\left(f_{i-1}, f_{i}\right)$, and $\rho^i_{eff} $ on the lhs of Eq. (\ref{lambda1}) are individually measured depending on the segment characteristic for each segment, the rhs is characterized by the source. So we can define this as a source-dependent quantity $\Lambda$. Then, for each segment,

\begin{equation}\label{lambda}
    \Lambda \equiv \left(\frac{8r_{0}}{D_{L}}\left(\frac{\mathcal{M}_{z}}{\mathcal{M}_{BNS}}\right)^{5/6} \right)^{-1}  \approx \frac{\Theta^i_{eff} \sqrt{\zeta^i\left(f_{i-1}, f_{i}\right)}}{\rho^i_{eff}}.
\end{equation}
Here, $0<\zeta^i\left(f_{i-1}, f_{i}\right) < 1$ and $0< \Theta^i_{eff} <6 $; therefore, $0 < \Lambda^i < 6/\rho^i_{eff}$.

In order constrain this quantity $\Lambda$, we now take into account the second set of observed data $D_2$ for the $i^{th}$ segment, $D_{2i} \equiv (\rho^i_{eff})$. Then the distribution of probability for $\Lambda$ in the $i^{th}$ segment can be written as

\begin{equation}\label{lambda_prob_seg}
    P(\Lambda | D_{1i}, D_{2i}, I) = \int d\Theta_{eff} P(\Lambda, \Theta_{eff}|D_{1i}, D_{2i}, I).
\end{equation}
The integrand can be expanded as

\begin{equation}
\begin{split}
    P(\Lambda, \Theta_{eff}|D_{1i}, D_{2i}, I) & = P(\Theta_{eff}|D_{1i}, D_{2i}, I)\\
    & \times P(\Lambda|\Theta_{eff}, D_{1i}, D_{2i}, I)\\
    & = P(\Theta_{eff}|D_{1i}, I)\\
    & \times P(\Lambda|\Theta_{eff}, D_{1i}, D_{2i}, I)
\end{split}
\end{equation}
where the probability $P(\Theta_{eff}|D_{1i}, I)$ is obtained in Eq.  (\ref{thetaeffprob_seg}) and

\begin{equation}
\begin{split}
    P(\Lambda|\Theta_{eff}, D_{1i}, D_{2i}, I) = \int & d\rho^i_{eff} P(\rho^i_{eff})\\
    & \times \delta\left(\Lambda - \Lambda(\Theta^i_{eff}/\rho^i_{eff})\right)
\end{split}
\end{equation}
where,

\begin{equation}\label{snr_eff_seg}
\begin{split}
      P(\rho^i_{eff})= \int & d\rho^i_1 \int  d\rho^i_2 \int d\rho^i_3 P(\rho^i_1)P(\rho^i_2)P(\rho^i_3) \\
      & \times \delta\left(\rho^i_{eff}-
      \sqrt{(\rho^i_1)^2+(\rho^i_2)^2+(\rho^i_3)^2}\right).
\end{split}
\end{equation}
Equations (\ref{fourangles_seg}) and (\ref{lambda_prob_seg}), give the probability distributions using information from the $i^{th}$ segment. Combining the information for all n segments we can write

\begin{equation}\label{fourangles_long}
    P(\Omega_{eff}|D_{1}, I) =   \frac{1}{N_{\Omega}}  \prod_{i=1}^{n} P(\Omega_{eff}|D_{1i}, I)
\end{equation}
and

\begin{equation}\label{lambda_prob}
    P(\Lambda|D_{1}, D_2, I) =  \frac{1}{N_{\Lambda}} \prod_{i=1}^{n}P(\Lambda | D_{1i}, D_{2i}, I)
\end{equation}
where $N_{\Omega}$ and $N_{\Lambda}$ are normalization constants and $D_2 \equiv (\rho^1_{eff}, \rho^2_{eff} \hdots \rho^n_{eff})$, for all the n segments.

We can now impose constraints on the chirp mass and redshift of the source. We use the measured value of the redshifted chirp mass ${\cal M}_z$, and we assume that the measured probability density of the quantity  $P({\cal M}_z)$ is Gaussian with the width of $\sigma_{{\cal M}_z}={\cal M}_z /\rho_{eff}$. Note that the justification of this assumption can be understood from Fig. 1 in \textcite{2021PhRvD.104d3014S}. We assume the error on $f_{max}$ to be small enough to neglect it. In addition to Eqs. (\ref{fmax}) and (\ref{snr_seg}), we use the constraint implied by the definition of the chirp mass:

\begin{equation} \label{massratio}
    \frac{{\mathcal{M}}}{M} = \left[\frac{q}{(1+q)^2} \right]^{3/5}< 4^{-3/5}.
\end{equation}
The last inequality in the above equation comes from the condition that $q_{max}=1$. Then the joint probability of ${\cal M}$ and $z$ given this information $I_Q$ is

\begin{equation}\label{qconstraint}
    P({\cal M}, z|I_Q, I) = P({\cal M}|I) P({z}|I) \frac{P(I_Q| {\cal M}, z,I)}{P(I_Q|I)}
\end{equation}
where

\begin{equation}\label{heaviside}
\begin{split}
    P(I_Q| {\cal M}, z,I) & = \mathcal{H} \left(\frac{\mathcal{M}}{2.8M_{\odot}}\frac{f_{max}(1+z)}{785 \rm {Hz}}\right)\\
   & \times \mathcal{H}\left(4^{-3/5} - \frac{\mathcal{M}}{2.8M_{\odot}}\frac{f_{max}(1+z)}{785 \rm {Hz}}\right)
\end{split}
\end{equation}
where ${\cal H}(x) $ is the Heaviside function.

We assume a flat cosmology with $\Omega_m=0.3$,  $\Omega_m + \Omega_\Lambda = 1, $ $\Omega_k = 0$, $H_0 = 67.3 \; \rm km s^{-1} \;\rm Mpc^{-1}$ \cite{2015PhRvD..92l3516A}. The relation between the luminosity distance $D_L$ and redshift $z$ is obtained from the analytic approximation given by \citet{Adachi2012}. The variation of comoving volume $V$ with redshift $z$ for this cosmology is

\begin{subequations}
\begin{equation}\label{dvdz}
  \frac{dV}{dz} = 4\pi D_H\frac{D_L^2}{(1+z)^2E_z},
\end{equation}
where,
\begin{equation}
D_H = c/H_0 \;  {\rm and}  \; E_z = \sqrt{\Omega_m(1+z)^3 + (1-\Omega_m)}.
\end{equation}
\end{subequations}
We assume flat priors on chirp mass $\mathcal{M}$ while the prior on redshift $z$ comes from Eq. (\ref{dvdz}) assuming that the rate density of mergers per unit comoving volume per unit time is constant in the Universe. Therefore, we have

\begin{equation}\label{chmprior_long}
    P({\cal M}|I) = \frac{1}{(\mathcal{M}_{max}- \mathcal{M}_{min})}
\end{equation}
and

\begin{equation}\label{zprior_long}
    P({z}|I) = \frac{dV}{dz}.   
\end{equation}
We can now include the information on the measurement of the redshifted chirp mass, and constraint on the quantity $\Lambda$ to impose further constraint on ${\cal M}$ and $z$. Taking into account the next observable $D_3 \equiv \mathcal{M}_z$, we can write

\begin{figure*}
\centering
\subfloat[ The probability distribution of neutron star masses mentioned in \protect\url{https://jantoniadis.wordpress.com/research/ns-masses/}. The green line is a summed Gaussian fit given in equation (\ref{m1_gaussfit}) for this distribution. The primary masses $m_1$ are drawn for using this fitting function. \label{fig:m1_fit}]{\includegraphics[width=\columnwidth]{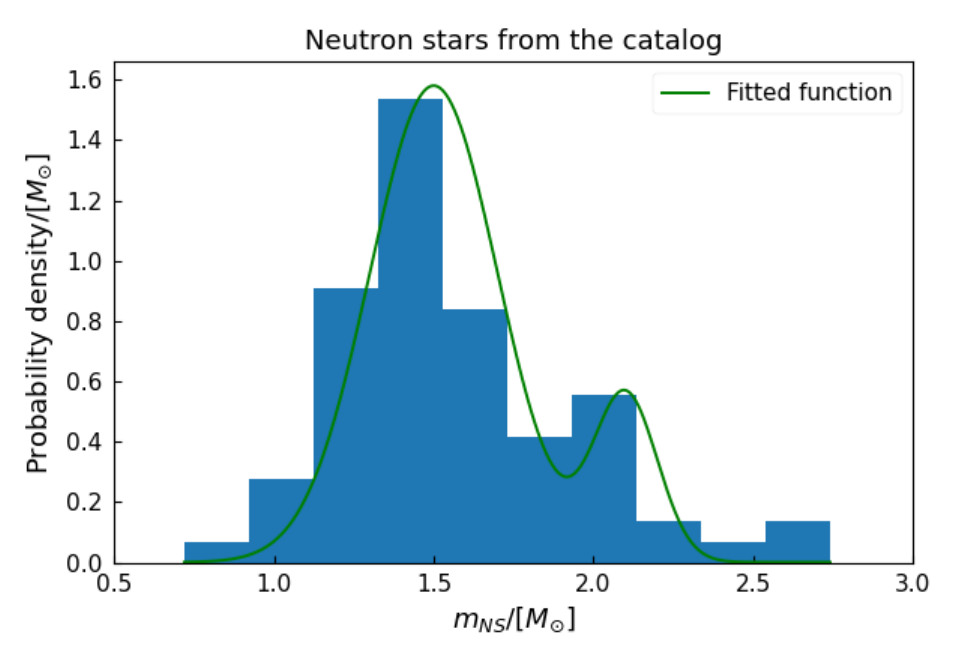}}\subfloat[ The probability distribution for $\mathcal{M}$ obtained for the low mass compact binary systems generated as mock sources using the distribution shown in Figure \ref{fig:m1_fit}. \label{fig:chm_fit}]{\includegraphics[width=\columnwidth]{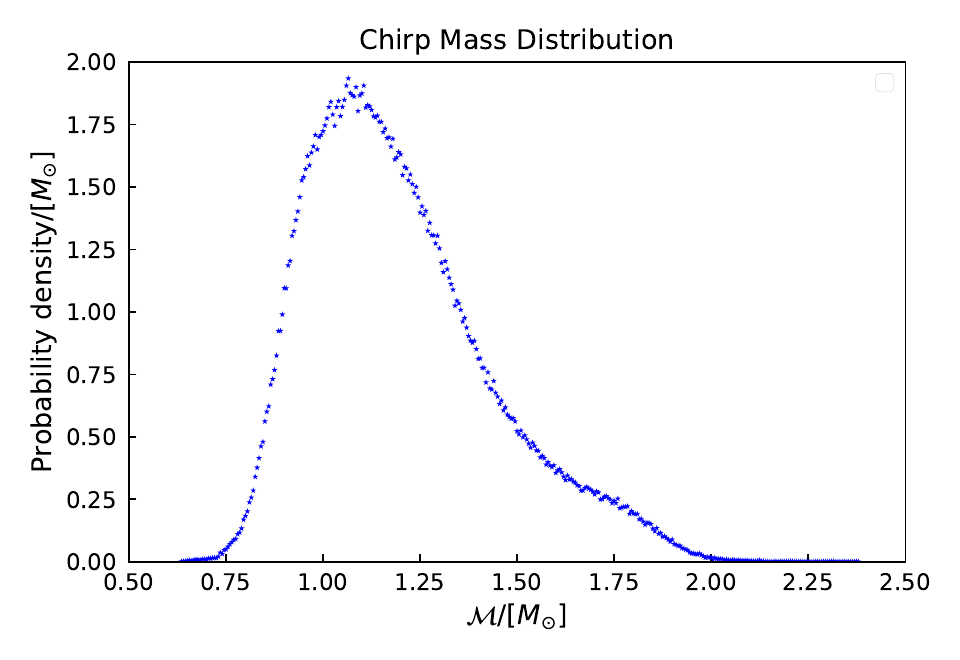}}
\caption{ The plots show the fit generated for the neutron star masses and the distribution of chirp masses of the sources in the mock catalog }
\end{figure*}

\begin{equation}\label{chirpzjointprob_long}
\begin{split}
   P(\mathcal{M}, z|D_1, D_2, D_3, I_Q, I) & \propto P(\mathcal{M}, z|I_Q, I) \int \int d\Lambda d \mathcal{M}_z\\
   & \times P(\Lambda| D_1, D_2, I)  P(\mathcal{M}_z) \\
   & \times \delta(\mathcal{M}_z - \mathcal{M}(1+z)) \\
   & \times \delta (\Lambda - \Lambda(\mathcal{M},z) )
\end{split}
\end{equation}
where we have used the form of $\Lambda$ from Eq. (\ref{lambda}) expressed in terms of $\mathcal{M}$ and $z$, given as

\begin{equation}
    \Lambda = \left(\frac{8r_{0}}{D_{L}(z)}\left(\frac{\mathcal{M}_{z}(\mathcal{M}, z)}{\mathcal{M}_{BNS}}\right)^{5/6} \right)^{-1}.
\end{equation}
Substituting the prior $P(\mathcal{M}, z|I_Q, I)$ required in the above Eq. (\ref{chirpzjointprob_long}) from Eq. (\ref{qconstraint}) as mentioned above, we obtain the joint probability of $\mathcal{M}$, and $z$.

Taking $D' \equiv (D_1, D_2, D_3)$ and $I' \equiv I_Q, I$, the marginalized distribution for $\mathcal{M}$ and  $z$ using Eq. (\ref{chirpzjointprob_long}) is given as

\begin{equation}\label{zprob_long}
    P(z|D', I') = \int d \mathcal{M} P(\mathcal{M}, z| D', I')
\end{equation}
and

\begin{equation}\label{chmprob_long}
    P(\mathcal{M}|D', I') = \int dz P(\mathcal{M}, z| D', I').
\end{equation}
The probability distribution for luminosity distance is obtained from the redshift using the cosmological assumptions described in Sec. \ref{sec:plan_long}

\begin{equation}\label{dlporb_long}
    \frac{dP}{dD_L} = \frac{dP}{dz}\frac{dz}{dD_L}
\end{equation}
Then we constrain the mass ratio using Eqs. (\ref{fmax}) and (\ref{massratio}) to express the mass ratio as a function of $\cal M$ and $z$:

\begin{equation}
     q({\cal M},z) = \frac{\xi}{2}-1-\sqrt{\frac{\xi^2}{4}-\xi}
\end{equation}
where

\begin{equation}
     \xi=\left( \frac{785{\rm Hz}}{f_{max}} \frac{2.8{\rm M}_\odot}{{\cal M}_z} \right)^{5/3}.
\end{equation}
Now we can write

\begin{equation}\label{qprob1_long}
    P(q|D',I') = \int\int d \mathcal{M} dz P(\mathcal{M}, z, q| D', I')
\end{equation}
We assume a flat prior on the mass ratio $P(q|I') = 1$, so

\begin{equation}\label{qprob_long}
    P(q|D',I') \propto \int \int d\mathcal{M} dz P(\mathcal{M}, z| D', I') \delta(q - q(\mathcal{M}, z)).
\end{equation}
Substituting Eq. (\ref{chirpzjointprob_long}) in Eq. (\ref{qprob_long}) gives the probability density for mass ratio $q$. The total mass of the binary $M$ is a obtained by using the observed value of the frequency $f_{max}$ and the probability distribution of redshift $z$. We can get the probability density of $M$  as

\begin{equation}\label{Mtotal_long}
    \frac{dP}{dM} = \int \frac{dP}{dz} \frac{dP}{d f_{max}} d z\, d f_{max}\delta(M-M(z,f_{max})) .
\end{equation}
Thus, using the observed SNRs, phases, redshifted chirp mass, and $f_{max}$, we have constrained the astrophysical parameters : chirp mass, redshift and the luminosity distance as well as total mass and mass ratio for a compact binary system.

\begin{table*}
\caption{ Details of two cases discussed in Sec. \ref{specific_cases}. The SNR values mentioned here are the accumulated SNRs in each of the three detectors. \label{tab:casedetail_long}}
\begin{ruledtabular}
\begin{tabular}{cccccccccccc}

 & $\rho_1$ & $\rho_2$ & $\rho_3$ & $\rho_{eff}$ & $\mathcal{M}$ & $M$ &  $z$ & $D_L$ & $q$ & $(\cos\delta, \alpha, \cos\iota, \psi)$ \\
 &  &  &  &  & ($M_{\odot}$) & ($M_{\odot}$) &  & (Gpc) &  & \\
\hline
Case 1 & 8.94 & 5.30 & 7.11 & 12.59 & 1.16 & 2.71 & 0.44 & 2.53 & 0.71 &  (-0.76, 3.72, -0.73, 5.90) \\
Case 2 & 18.16 & 24.70 & 41.56 & 51.64 & 1.68 & 3.91 & 0.1 & 0.47 & 0.74 & (0.07, 1.29,  -0.59,  4.35) \\
\end{tabular}
\end{ruledtabular}
\end{table*}

\begin{table*}
\caption{ Errors on the recovered values for the two cases mentioned in Table \ref{tab:casedetail_long}. The error on the parameters is the estimate for the spread of $90\%$ probability about the median for the respective parameters. The area estimated for the localization of declination and right ascension, $(\delta, \alpha)$  is the spread for $90\%$ probability about the peak value of the $(\delta, \alpha)$ distribution. \label{tab:errordetail_long}}
\begin{ruledtabular}
\begin{tabular}{cccccccc}
&  $\Delta \mathcal{M} $ & $\Delta M$ & $\Delta z$ & $\Delta D_L$ &
$\Delta q$  & Area for $(\delta, \alpha)$ localization ($A_{90}$)\\
 & ($M_{\odot}$) & ($M_{\odot}$) &  & (Gpc) &  & (sq deg)\\ 
\hline
Case 1  & 0.35 & 0.69 & 0.36 & 2.49 & 0.55 & $ 2.02\times 10^4 $ \\
Case 2  & 0.13 & 0.16 & 0.04 & 0.20 & 0.37  &  38.41  \\
\end{tabular}
\end{ruledtabular}
\end{table*}

\begin{table*}
\caption{ The assumed coordinates of the sites for the ET in this analysis. $\gamma$ is measured counterclockwise from East to the bisector of the interferometer arms, and $\eta$ is the angle between the interferometer arms. \label{tab:site_coordinates}}
\begin{ruledtabular}
\begin{tabular}{ccccc}
Site for ET & Latitude  & Longitude & Orientation & Arm angle  \\
& $\lambda$ & L & $\gamma$ & $\eta $  \\ 
\hline
Virgo & 43.68 \textdegree & 10.49 \textdegree & 84.84 \textdegree & 60 \textdegree \\ 
Sardinia & 40.12 \textdegree & 9.01 \textdegree & 0 \textdegree & 60 \textdegree \\ 
Maastricht & 50.85 \textdegree & 5.69 \textdegree & 0 \textdegree & 60 \textdegree \\ 
\end{tabular}
\end{ruledtabular}
\end{table*}

\section{Mock Source Catalog}\label{sec:injections_bns}

We generate a mock population of low mass compact binaries for this analysis assuming that the distributions of masses, distances, locations in the sky, and polarizations are independent. The cosmological assumption are as explained in the previous section. We assume that the mass distribution is the same for all distances.

Such a population of compact binary sources, distributed uniformly in comoving volume, is not a realistic one since it does not take into account the dependence of the star formation rate on redshift and also neglects the delays between formation and coalescence. The calculation using this assumption of sources distributed uniformly in comoving volume, is done for illustrative purposes so as to access the quality of the method employed for data analysis. We choose this simplified assumption for creating the mock population of low mass compact binary sources to study the biases if any, seen in the recovery of the parameters and to get an estimate of the accuracy of the recovered parameters. We discuss the analysis of more realistic populations in the next paper in this series \cite{2022A&A...667A...2S}.

The probability distributions of angular distribution $\cos\delta, \alpha/ \pi$, $\cos \iota$ and $\psi/ \pi$ are assumed to be uncorrelated and are distributed uniformly over the range $[-1,1]$. We generate sources spread in redshift range $[0.01, 1.0]$, assuming  that the rate density of mergers per unit comoving volume per unit time is constant in the Universe.

ET will be able to detect BNS binaries of total masses $\sim 3 \; M_{\odot}$ up to $z \sim (2-3)$  \cite{Sathyaprakash:2019nnu}, but we restrict the sources to redshift $z < 1$ so as to have substantial number of sources crossing the detection threshold we set for the analysis. 

The SNR for a given binary source in the mock population is calculated using the PSD $S_{h}\left(f\right)$, for the ET-D noise curve \cite{2011CQGra..28i4013H} obtained from the ET web-page \url{http://www.et-gw.eu/index.php/etsensitivities}. We choose the following criterion to define a detection: a threshold value of accumulated effective SNR $\rho_{eff}>8$ and the SNR for $i^{th}$ segment in the $j^{th}$ detector $\rho^i_j> 3$ in at least one segment, for $j = (1,2,3)$ corresponding to the three ET detectors comprising single ET. The justification of choosing only the segments with $\rho^i_j> 3$ and a comparison of detection thresholds is described in the Appendix.

To get a mass distribution for the low mass binary mock population, we use a summed Gaussian fit to the masses in the catalog \cite{2012ARNPS..62..485L} to get the distribution of primary mass $m_1$. The fitting function is

\begin{equation}\label{m1_gaussfit}
    p(m_1) \propto \left( \mathcal{N}(1.5, 0.2^2) + 0.35 \times \mathcal{N}(2.1, 0.1^2) \right)
\end{equation}
where $\mathcal{N}(\mu, \sigma^2)$ denotes the normal distribution with the mean $\mu$ and standard deviation $\sigma$. The fit is shown in Fig. \ref{fig:m1_fit}. The minimum and maximum mass limits for the binary components are restricted to $m_1, m_2 > m^{NS}_{min}= 0.72 \; M_{\odot}$ and $m_1, m_2 < 2.74 \; M_{\odot}$ in agreement with the values in the catalog. The mass ratio $q = m_2/m_1$ is chosen uniformly from the range [$m^{NS}_{min}/m_1 $,1]. The values of $m_1$ and $m_2$ then give the distribution of chirp mass $\mathcal{M}$ of the mock population. This distribution of $\mathcal{M}$ in this population is shown in Fig. \ref{fig:chm_fit}.

\begin{figure*}
\subfloat[\label{fig:loc1_final_thetaphi}]{\includegraphics[width=0.5\textwidth]{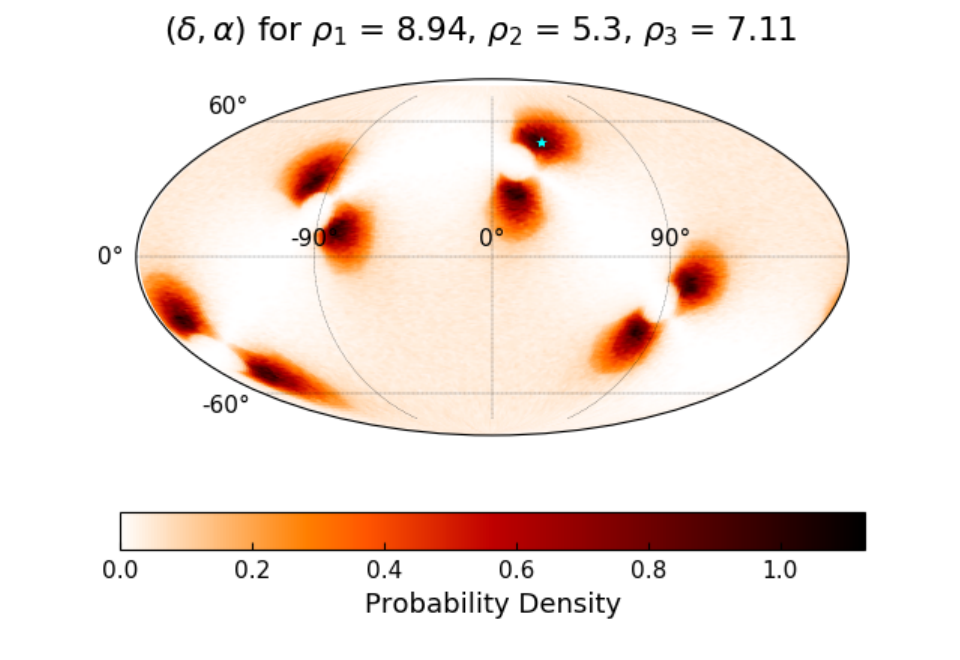}}
\subfloat[\label{fig:loc1_final_incpsi}]{\includegraphics[width=\columnwidth]{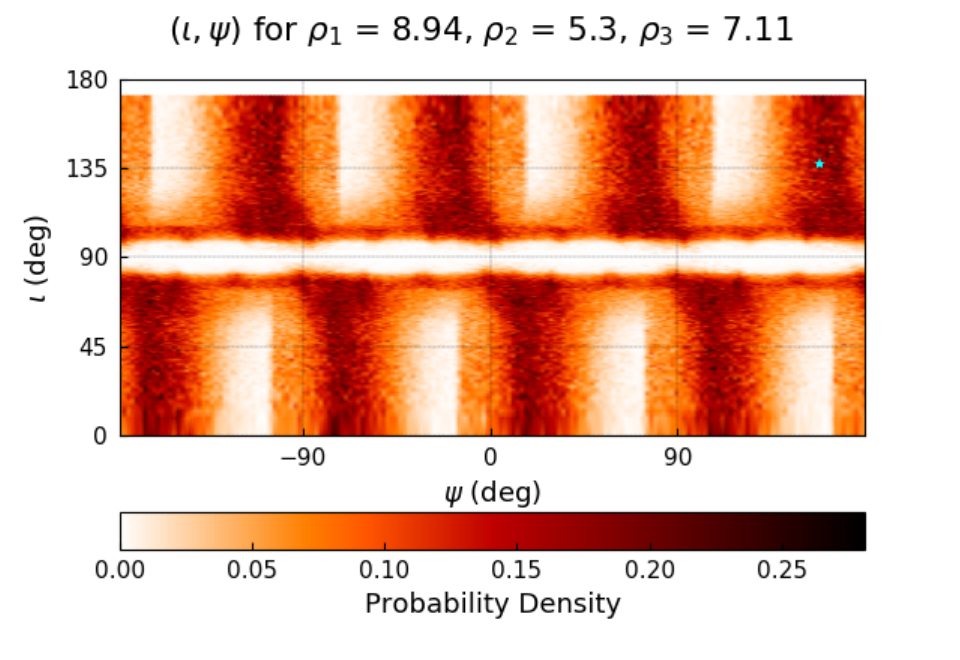}}
\caption{ The final probability distribution for $(\delta, \alpha)$ (left) and $(\iota, \psi)$ (right)  obtained for Case 1. The signal is detectable in the band for only one segment for a duration of 1.38 minutes. The blue star in the plots denotes the actual source parameters.}
\label{fig:loc_final_case1}
\end{figure*}

\section{Results}

In our analysis, we consider only the inspiral part of the signal from coalescing low mass binaries. The sources which cross the detection threshold are taken up for the analysis. These are referred to as the detected \emph{sources} in the discussion hereafter, and their parameters are denoted by subscript '\emph{s}'. The signal from each detection is split into 5-min segments as described in \S \ref{sec:plan_long}, assuming that the change in the antenna response functions is negligible during that period. For each segment, we generate a four-dimensional space of $\cos\delta, \alpha/ \pi, \cos \iota$, and $\psi/ \pi$, randomly distributed uniformly over the range $[-1,1]$. This 4D grid is constrained using Eqs. (\ref{theta-ratios}) and (\ref{phi-diff}) to get a distribution of $\Omega_{eff}$ from Eq. (\ref{fourangles_seg}). This gives us the sky localization of declination and right ascension $(\delta, \alpha)$ on the celestial sphere, and constraint on the source angles $(\iota, \psi)$. 
Since we assumed that the antenna response functions remain unchanged during 5 min, this introduces a systematic error in the location which is smaller than $1.25^{\circ}$. In order to avoid dealing with this potential contribution to the localization error we keep the bin size of  the sphere  to be $1.28$  square degrees.

We get the posterior distribution of $\Theta^i_{eff}$ from Eq. (\ref{thetaeffprob_seg}) for each segment using the information obtained for $\Omega_{eff}$. Since we assume that the effective SNR $\rho^i_{eff}$ is known for each segment in the three ET detectors, we use this information about the measured values of $\rho^i_{eff}$ and the distribution of $\Theta^i_{eff}$ obtained in each segment to provide a constraint on $\Lambda$, given by Eq. (\ref{lambda_prob_seg}). The function $\Theta^i_{eff}$ varies for each segment as the antenna response function changes with the rotation of Earth. $\zeta^i\left(f_{i-1}, f_{i}\right)$ varies due to change in the limit of integration as specified in Eq. (\ref{zetafunc_seg}). From Eq. (\ref{lambda}), we see that $\Lambda$, is a source-dependent quantity. This process is repeated for all the segments and the final distribution for $\Omega_{eff}$ and $\Lambda$  is obtained by combining information obtained from all the segments as given by Eqs. (\ref{fourangles_long}) and (\ref{lambda_prob}).

In order to estimate the other parameters of the binary system, we construct a 2D grid for redshift $z$ and chirp mass $\mathcal{M}$, with ranges from $0.005 \leq z \leq 1.5$ and $ 0.6 \leq \mathcal{M}/M_{\odot} \leq 2.5$. The limits of these ranges extend beyond the limits of the mock sources so that the sources at the edge of the mock population can be recovered correctly. The prior on $z$ is given by Eq. (\ref{zprior_long}) with details given in Sec. \ref{sec:plan_long}. The prior on $\mathcal{M}$ is assumed to be flat, given by Eq.  \ref{chmprior_long}).

Since we assume that the redshifted chirp mass is known from match filtering, we use this information to select the appropriate values from the 2D  ($z,\mathcal{M}$) grid. The observed value of $f_{max}$, the frequency at the end of the inspiral, gives the information about the mass ratio $q$ which further limits the valid grid points. Last, since we have a distribution of $\Lambda$, obtained by combining information from all the detected segments, we use this distribution to get the joint probability of $(z, \mathcal{M})$ using Eq.  (\ref{chirpzjointprob_long}). The distributions for luminosity distance $D_L$ and mass ratio $q$ are obtained from Eqs. (\ref{dlporb_long}) and (\ref{qprob_long}), respectively, and the distribution for total mass $M$ is obtained using Eq. (\ref{Mtotal_long}). The subscript '\emph{median}' represents the median value of these distributions obtained for the parameters of the binary system.

\begin{figure*}
\subfloat[\label{fig:loc2_a}]{\includegraphics[width=\columnwidth]{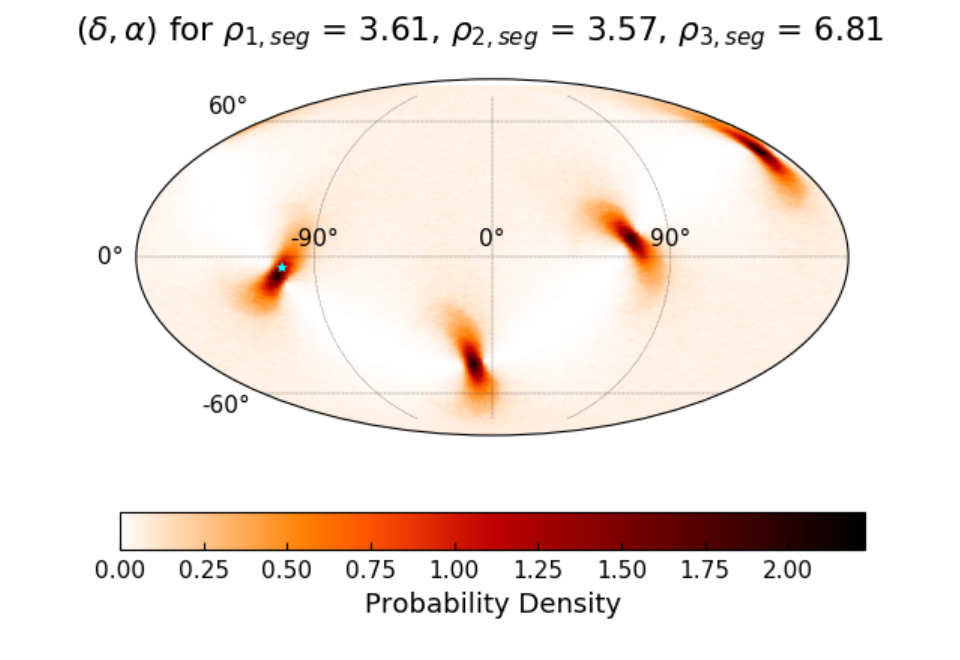}}
\subfloat[\label{fig:loc2_b}]{\includegraphics[width=\columnwidth]{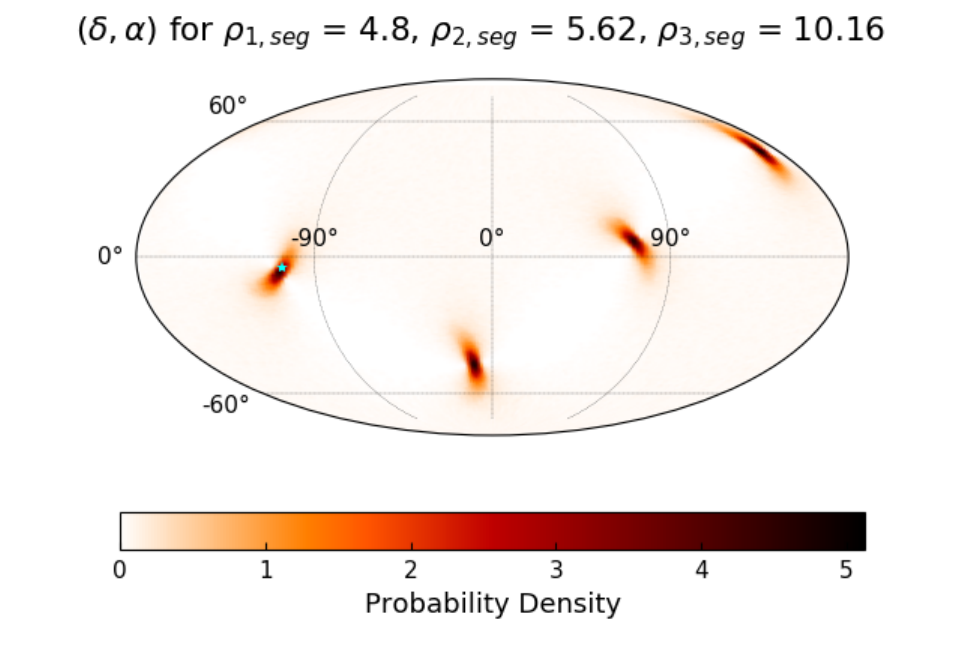}}\\
\subfloat[\label{fig:loc2_c}]{\includegraphics[width=\columnwidth]{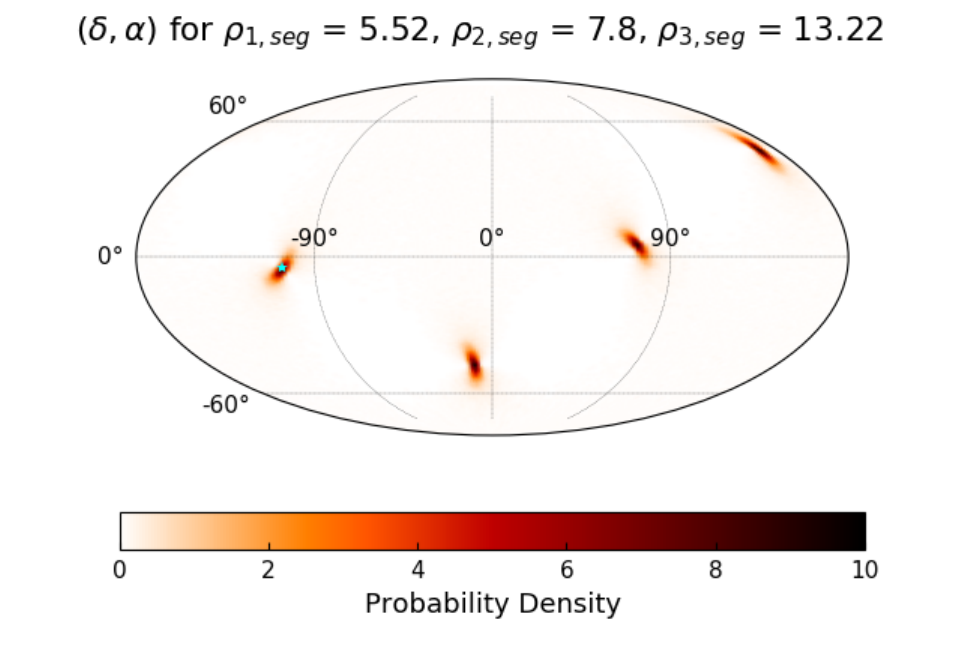}}
\subfloat[\label{fig:loc2_d}]{\includegraphics[width=\columnwidth]{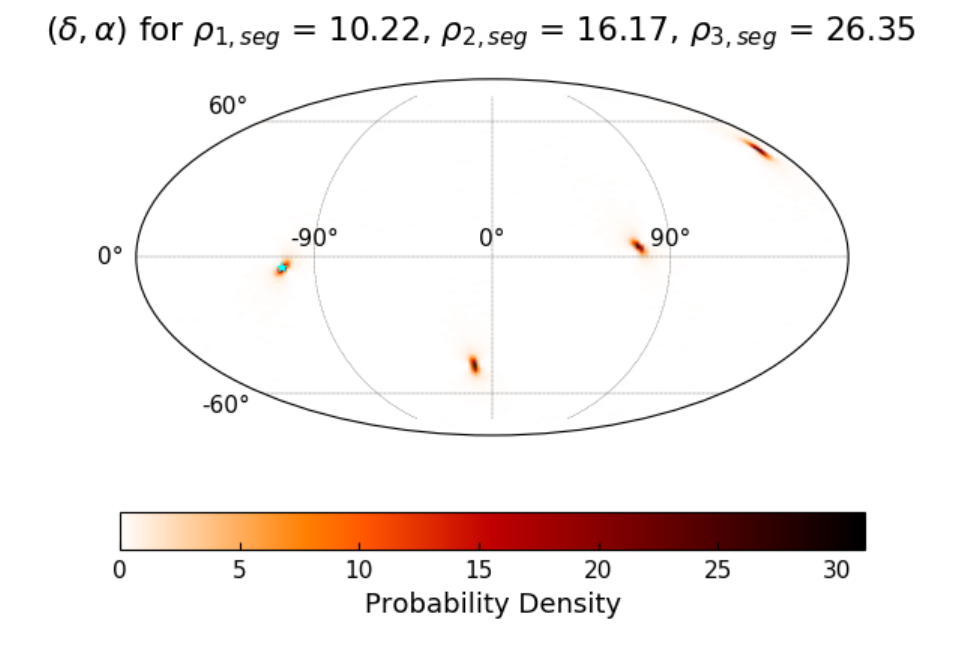}}
\caption{ Localization in each segment for case 2. Plots shown here are for the (a) first, (b) third, (c) fifth, and (d) last seventh segment from the time the signal crosses the threshold of detection up till the last segment. There are seven segments as the signal is 30.63 min long in this case. Going from a $\rightarrow$ b $\rightarrow$ c $\rightarrow$ d, shows the distributions obtained successively in time for the declination and right ascension $(\delta, \alpha)$. The SNR values $\rho_{j,seg}$ denote the SNR generated in the particular segment of the $j^{th}$ detector. The blue star denote the actual value of declination and right ascension $(\delta, \alpha)$ of the detected source on the celestial sphere.}
\label{fig:loc2seg}
\end{figure*}

\begin{figure*}
\subfloat[\label{fig:loc2_final_thetaphi}]{\includegraphics[width=\columnwidth]{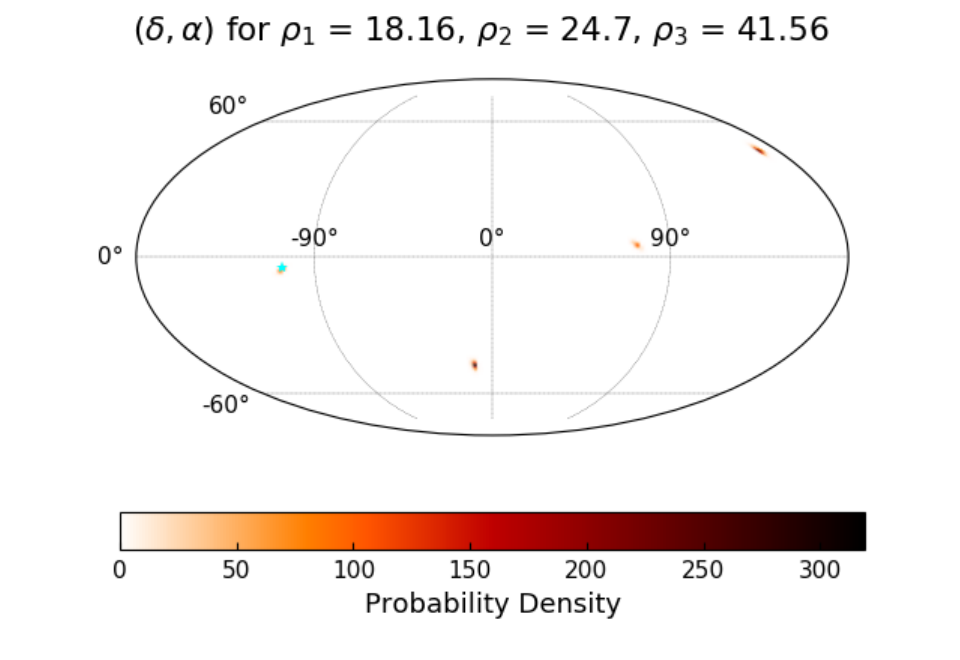}}
\subfloat[\label{fig:loc2_final_incpsi}]{\includegraphics[width=\columnwidth]{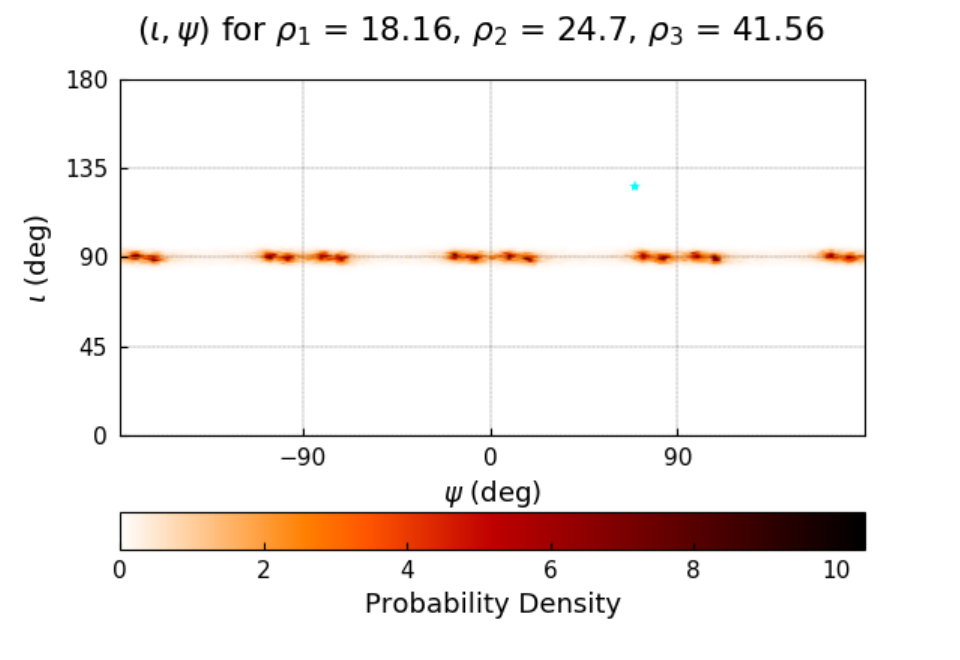}}
\caption{ The left plot shows the final probability distribution for $(\delta, \alpha)$ obtained for case 2 by combining probabilities for all the segments (some of which are shown in Fig. \ref{fig:loc2seg}). The corresponding final probability of $(\iota, \psi)$ is shown in the right plot. The SNR values denote the accumulated SNRs in each detector. The blue star denote the declination and right ascension $(\delta, \alpha)$ value of the detected source on the celestial sphere. }
\label{fig:loc_final_case2}
\end{figure*}

\subsection{Analysis of particular cases}\label{specific_cases}

To demonstrate the method described above, we discuss two cases. The details of the two cases are mentioned in Table \ref{tab:casedetail_long}. In case 1, we consider a low mass compact binary system with chirp mass $\mathcal{M} = 1.16 M_{\odot}$, total mass $M = 2.71 M_{\odot} $ located at redshift $z = 0.44$. The inspiral signal from this binary enters the ET band at 1Hz but is detected only at the start of $924^{th}$ segment. In this segment,  $(f_{i-1}, f_i) \sim (20.95, 1128.07)$ Hz. The SNR values in this segment cross the detection threshold with $\rho_{1,seg} = 8.94, \; \rho_{2, seg} = 5.30, \; \rho_{3, seg} = 7.11 $. Since this is the last segment of the inspiral signal for this binary, we have the information from this one segment only which is 1.38 min long. The distributions for declination, right ascension $(\delta, \alpha)$ and  inclination, polarization angle $(\iota, \psi)$ are shown in Fig. \ref{fig:loc_final_case1}. In the distribution for $(\delta, \alpha)$,  we see the degeneracy in the recovered angles coming from the nature of dependence of $\Theta$ function on the angles. In particular, there is the symmetry about the equatorial plane i.e ($\theta \rightarrow -\theta$), and also with respect to the rotation by $90^{\circ}$ about the longitude i.e ($\phi \rightarrow \phi + 90^{\circ} $), where $(\theta, \phi)$ are the sky coordinates in the detector frame (see Fig. 2 in  Ref. \cite{2021PhRvD.104d3014S} for a better understanding). This results in eight images of possible locations of the source.  The plot shown in Fig. \ref{fig:loc1_final_thetaphi} shows the localization of $(\delta, \alpha)$ is fairly well constrained while the constraint on $(\iota, \psi)$ as seen in Fig. \ref{fig:loc1_final_incpsi} is much weaker. The area of $90\%$ probability ($A_{90}$) about the peak for declination and right ascension $(\delta, \alpha)$ in this case is $ 2.02 \times 10^4 $ square degrees.

Case 2 is a detected source with $\mathcal{M} = 1.68 M_{\odot}$, total mass $M = 3.91 M_{\odot}$ located at redshift $z = 0.1$. After entering the ET band at 1Hz, the signal is detected in the $775^{th}$ segment. In this first detected segment $\rho_{1, seg} = 3.61, \; \rho_{2, seg} = 3.57, \; \rho_{3, seg} = 6.81$ in the three ET detectors and  $(f_{i-1}, f_i) \sim (6.16, 6.58)$ Hz. For this case, the inspiral signal stays for 30.63 minutes in the detectable range of ET, with a total of seven segments. In the last segment $(f_{i-1}, f_i) \sim (26.47, 1023.35)$Hz. The plots in Fig \ref{fig:loc2seg} show first, third, fifth and the final seventh segment from the time the signal crosses the threshold of detection. Moving from  \ref{fig:loc2_a} $\rightarrow$ \ref{fig:loc2_b} $\rightarrow$ \ref{fig:loc2_c} $\rightarrow$ \ref{fig:loc2_d}, we can see the reduction in the area of localization in each segment with an increase in $\rho_{eff}$ for each segment. We combine the probabilities from all these segments to get the final probability shown in Fig. \ref{fig:loc_final_case2}. The main advantage of combining the information from all the segments in this manner is the breaking of the angular degeneracy. Now, instead of a large spread of possible localization of the source, we have a much smaller region left, spread over only four images. The area of $90\%$ probability about the peak for declination and right ascension $(\delta, \alpha)$ in this case is 38.41 square degrees.

We use the observed value of redshifted chirp mass $\mathcal{M}_z$ to limit the initial 2D grid of redshift $z$ and chirp mass $\mathcal{M}$ to those which satisfy this observed value within the error of measurement $\sigma_{\mathcal{M}_z}$. The observable $f_{max}$, the frequency at the end of the inspiral, gives the information about the mass ratio $q$ restricting the grid even more using Eq. (\ref{qconstraint}). We get the final distribution for the joint probability distribution for $z$ and  $\mathcal{M}$, using the distribution for $\Lambda$ from Eq. (\ref{chirpzjointprob_long}).

\begin{figure*}
\subfloat[\label{fig:case1_zchm}]{\includegraphics[width=\columnwidth]{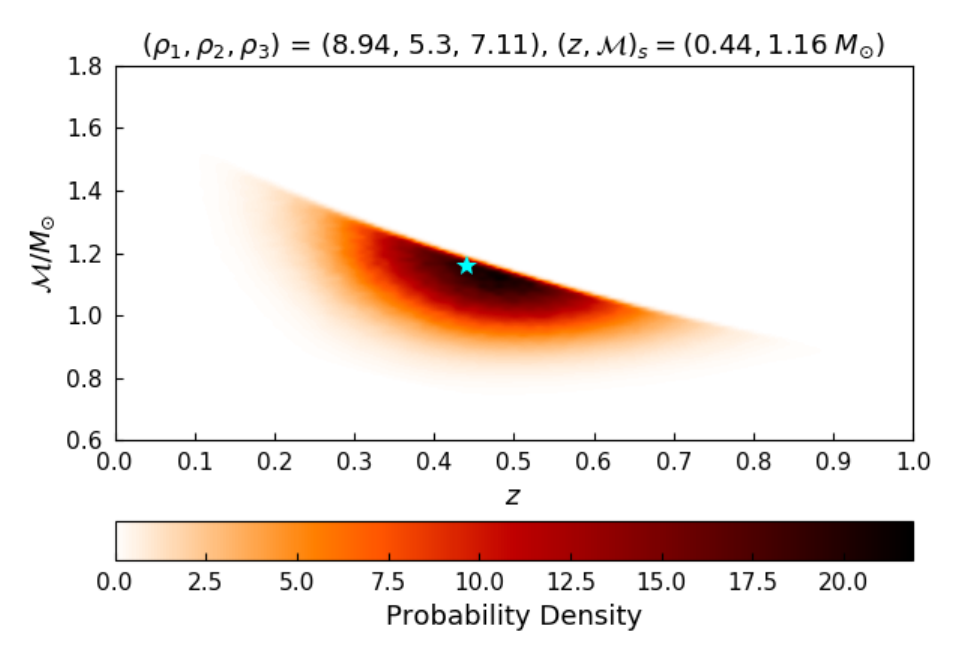}}
\subfloat[\label{fig:case2_zchm}]{\includegraphics[width=\columnwidth]{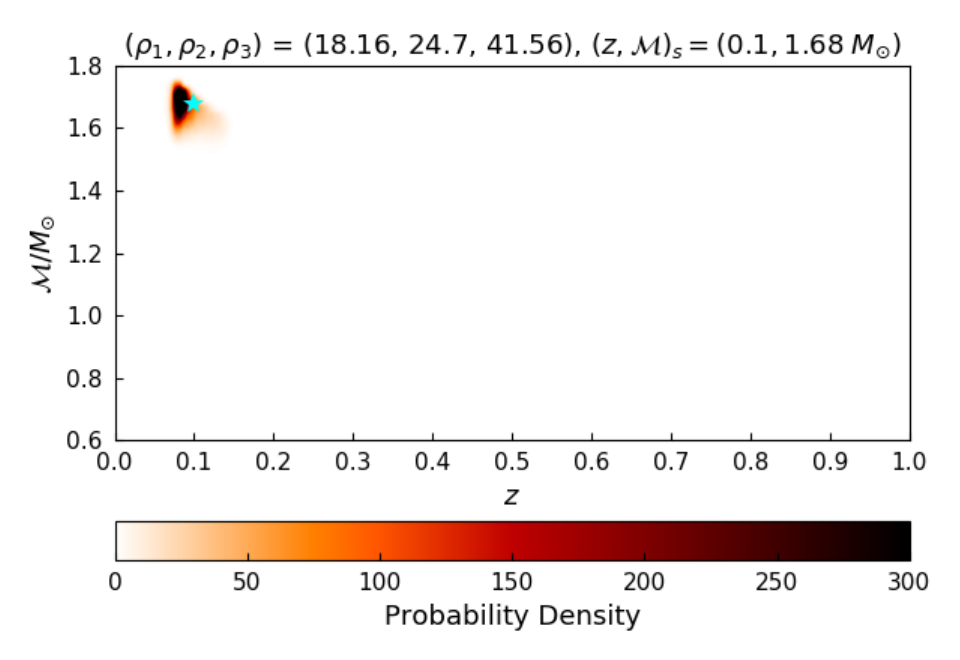}}\\
\subfloat[\label{fig:case1_z}]{\includegraphics[width=\columnwidth]{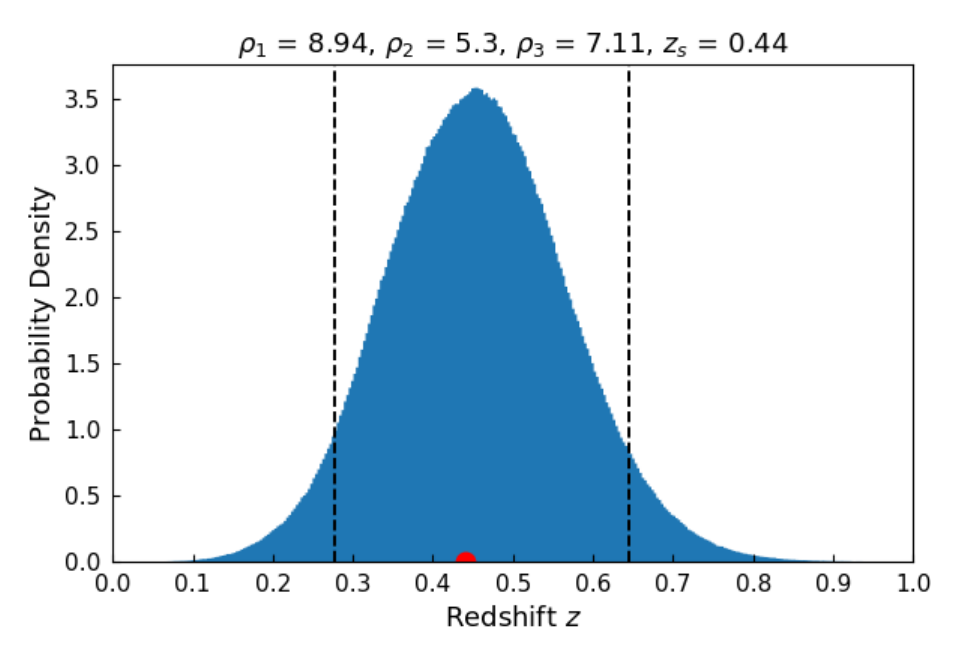}}
\subfloat[\label{fig:case2_z}]{\includegraphics[width=\columnwidth]{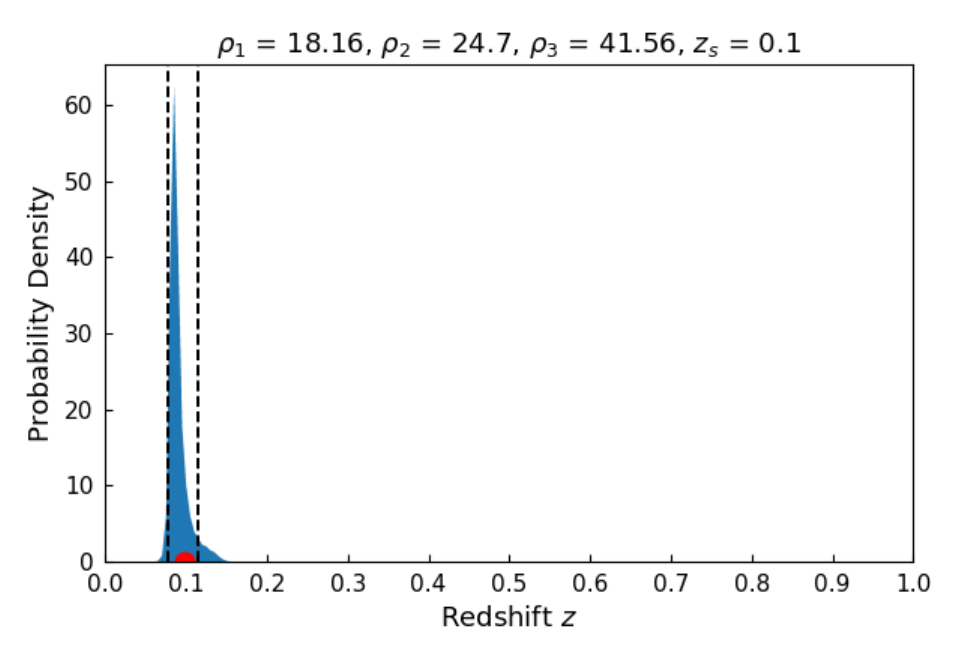}}\\
\subfloat[\label{fig:case1_chm}]{\includegraphics[width=\columnwidth]{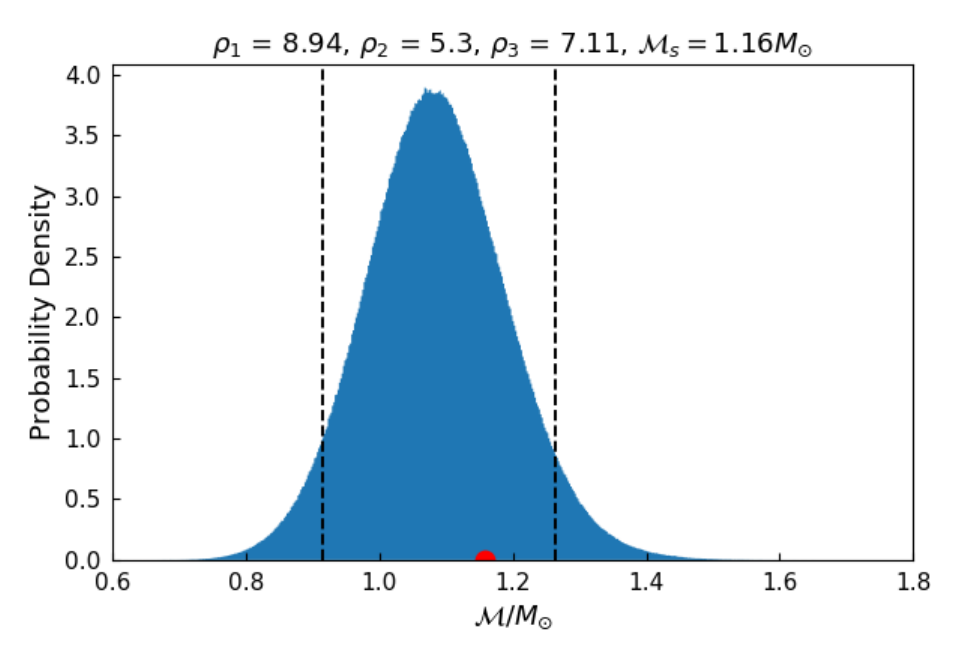}}
\subfloat[\label{fig:case2_chm}]{\includegraphics[width=\columnwidth]{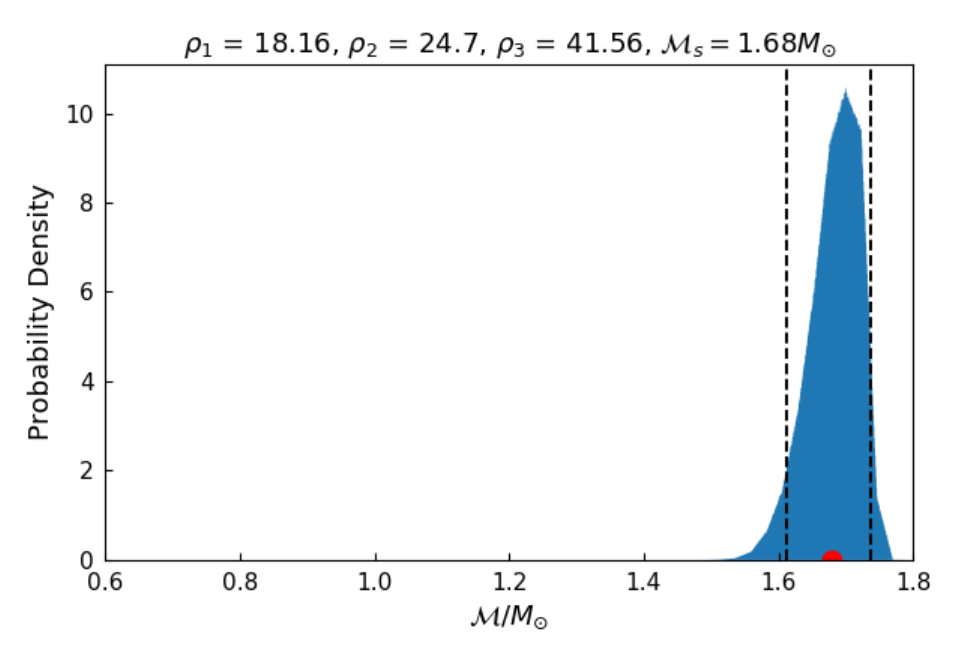}}
\caption{ The distributions obtained for $\mathcal{M}$ and $z$ for case 1 (left) and case 2 (right). The top panel shows the joint probability of $\mathcal{M}$ and $z$. The marginalized probabilities of these quantities are shown in the lower panels. The blue star in the top panel plots and red dot in the middle and bottom panel plots denote the actual source value of the parameters.}
\label{plots_z_chm}
\end{figure*}

\begin{figure*}
\subfloat[\label{fig:case1_dl}]{\includegraphics[width=\columnwidth]{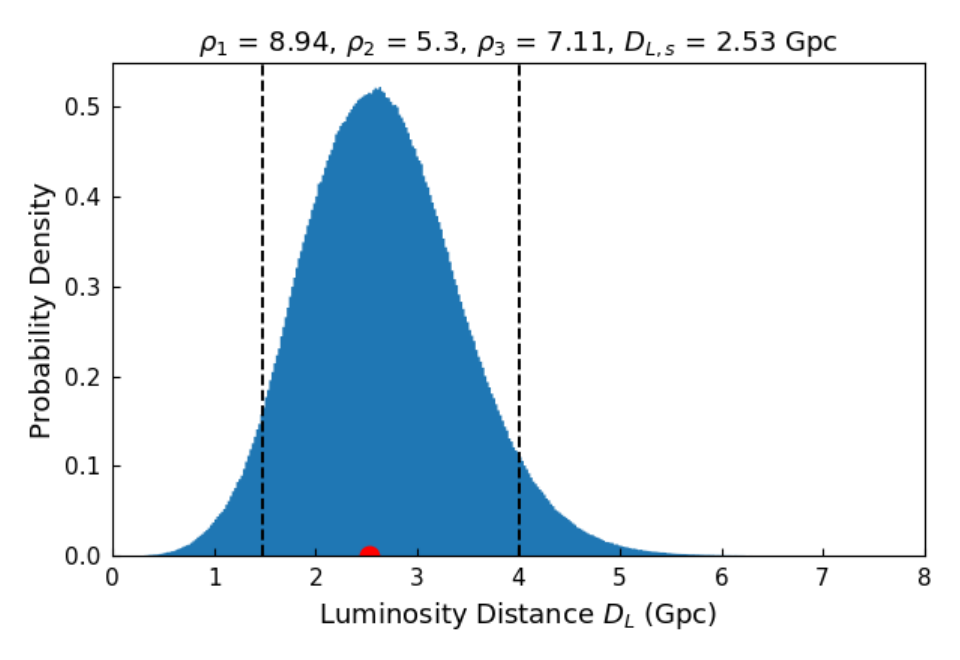}}
\subfloat[\label{fig:case2_dl}]{\includegraphics[width=\columnwidth]{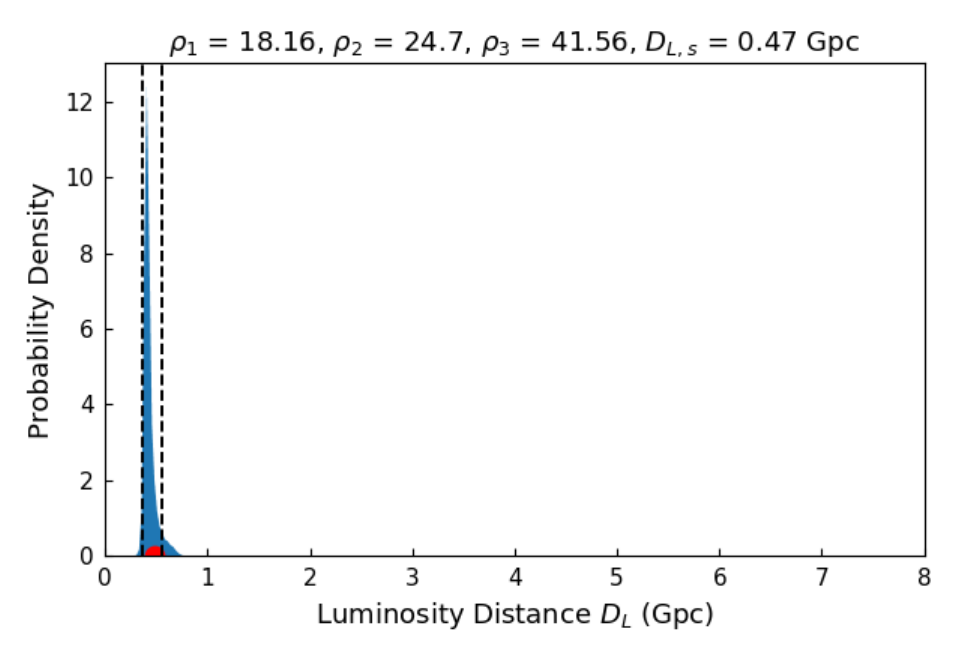}}\\
\subfloat[\label{fig:case1_q}]{\includegraphics[width=\columnwidth]{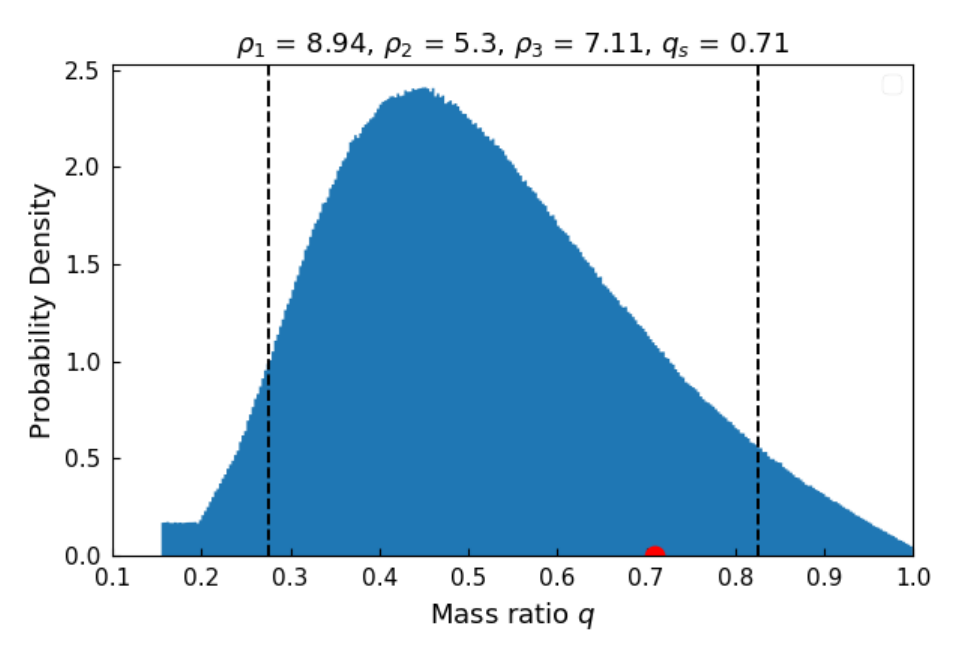}}
\subfloat[\label{fig:case2_q}]{\includegraphics[width=\columnwidth]{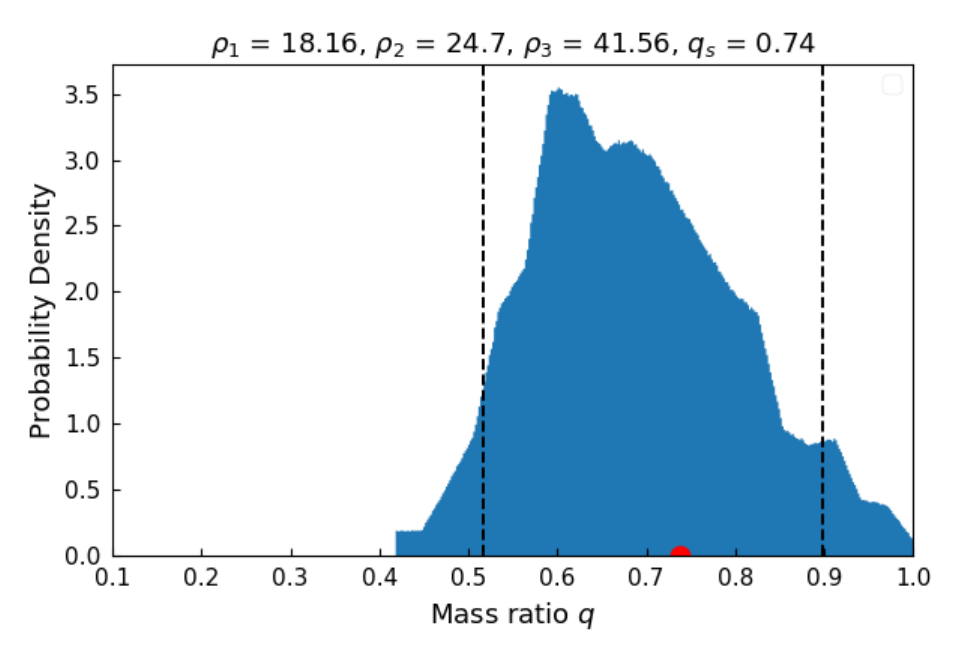}}\\
\subfloat[\label{fig:case1_M}]{\includegraphics[width=\columnwidth]{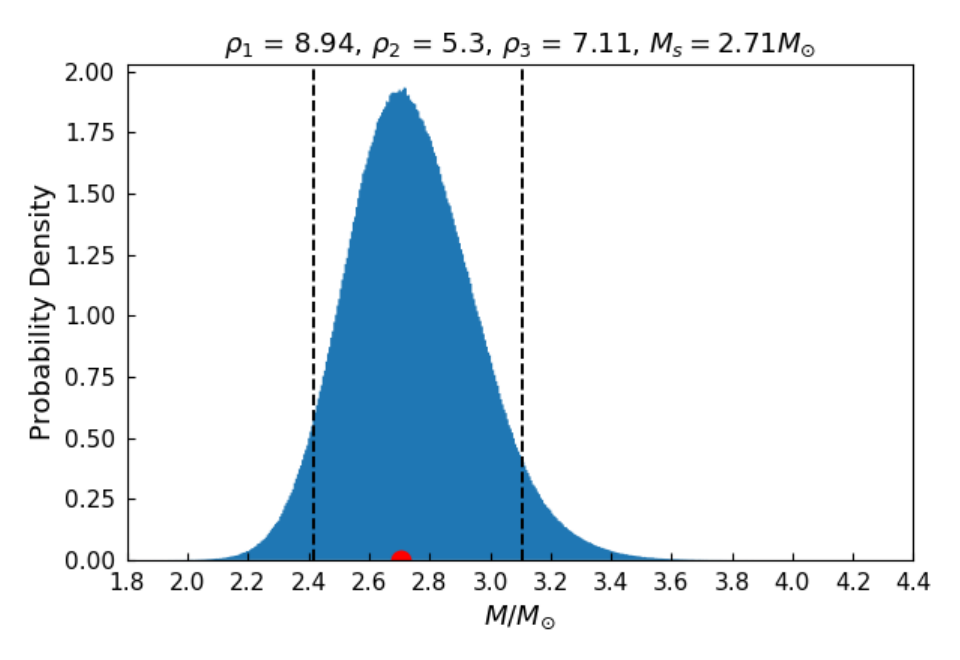}}
\subfloat[\label{fig:case2_M}]{\includegraphics[width=\columnwidth]{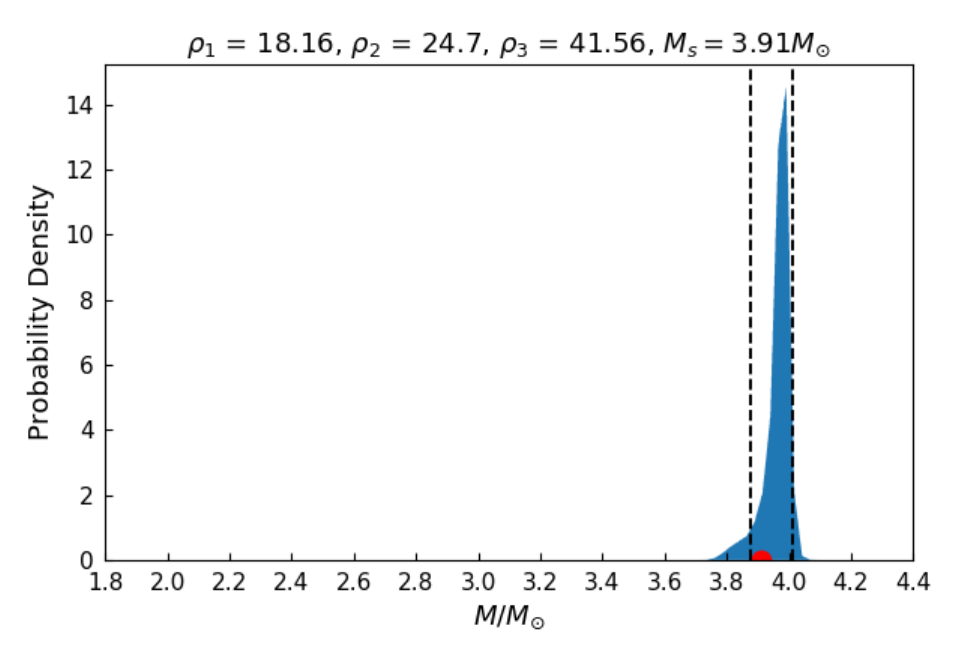}}
\caption{ The distribution of luminosity distance $D_L$ (a,b), mass ratio $q$ (c,d), and total mass $M$ (e,f), for case 1 (left) and case 2 (right). The red dot denotes the actual source value of the parameter.}
\label{fig:plots_dl_q_M}
\end{figure*}

The distributions obtained for $\mathcal{M}$ and $z$ for case 1 and case 2 are shown in left and right panels of Fig. \ref{plots_z_chm} respectively. Figures \ref{fig:case1_zchm} and \ref{fig:case2_zchm} are the joint probabilities for $\mathcal{M}$ and $z$ obtained using equation (\ref{chirpzjointprob_long}). The Figures \ref{fig:case1_z} and \ref{fig:case2_z} show the marginalized probability distribution recovered for the redshift $z$. The 90\% error about the median of the recovered redshift decreases from 0.36 for case 1 to 0.04 for case 2. Figures \ref{fig:case1_chm} and \ref{fig:case2_chm} show the marginalized probability distribution for $\mathcal{M}$ for which the error about the median reduces from $0.35 M_{\odot}$ to $0.13 M_{\odot}$. Figure \ref{fig:plots_dl_q_M} shows the corresponding distributions of luminosity distance $D_L$ and mass ratio $q$ and total mass $M$ for case 1 and case 2. The values of the error estimates for all the parameters are mentioned in Table \ref{tab:errordetail_long}.

\subsection{Analysis of mock population}

We generate a 2D grid of 80 000 points for ($m_1, q$) using Eq. (\ref{m1_gaussfit}) and choosing the mass ratio to be uniformly distributed in the range [$m^{NS}_{min}/m_1 $,1]. We then apply the limits on $m_1, m_2$ as discussed in Sec. \ref{sec:injections_bns}. After applying these cuts, we are left with 42 514 binary sources which are then randomly distributed in redshift using Eq. (\ref{zprior_long}). Each compact binary system is then assigned a set of random values of the four angles: angle of declination $\delta$, right ascension $\alpha$, polarization angle $\psi$ and inclination angle $\iota$, of the binary with respect to the direction of observation.

Out of these 42 514 sources, 17 994 compact binary systems cross the detection threshold and are taken up for the analysis. We discuss the localization and constraints on the binary parameters in the following sections.

\begin{figure}[ht]
\centering
\includegraphics[width = \columnwidth]{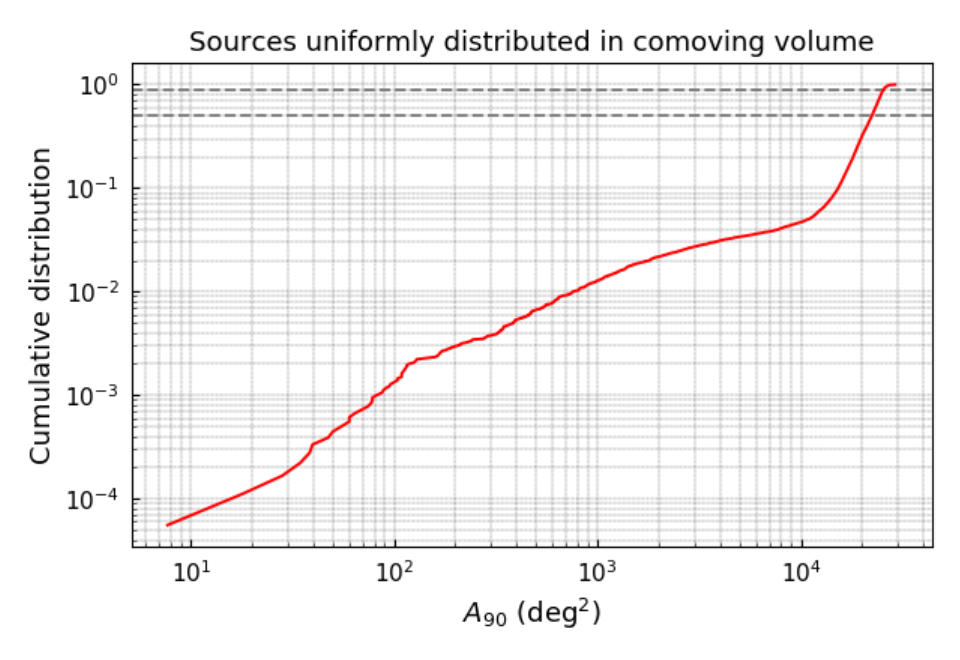}
\caption{ Localization of the sources distributed uniformly in comoving volume  using single ET. The figure shows the cumulative distribution of $90\%$ probability about the peak for $(\delta, \alpha)$. The two dashed lines denote the  50\% and 90\% cumulative probability. \label{fig:area_cummu}}
\end{figure}

\subsubsection{Localization capability}\label{loc_cap}

The distribution of sky localization using a single ET is shown in Fig. \ref{fig:area_cummu}. The sources are assumed to be distributed uniformly in comoving volume. It shows the cumulative distribution of area of $90\%$ probability, $A_{90}$ of the sky localization. The two dashed lines in this figure denote the 50\% and 90\% cumulative probability. 100\% of the analyzed sources distributed uniformly in comoving volume, are localized within $\sim 2.90 \times 10^4$ square degrees which is $\sim 70\%$ of the whole sky. 90\% of the analyzed sources are localized within $\sim 2.55 \times 10^4$  square degrees and 50\% are within $\sim 2.23 \times 10^4$ square degrees. For the best case, we see that using this method of analyzing the long-duration signal, single ET in triangular configuration can constrain the localization area for 90\% probability region of $(\delta, \alpha)$ to a minimum value of 7.68 square degrees, for $\rho_{eff} = 184.59$, but only $\approx 1\%$ of binaries can be localized within 800 square degrees.

\begin{figure*}
\subfloat[\label{fig:area_rho_inc}]{\includegraphics[width=\columnwidth]{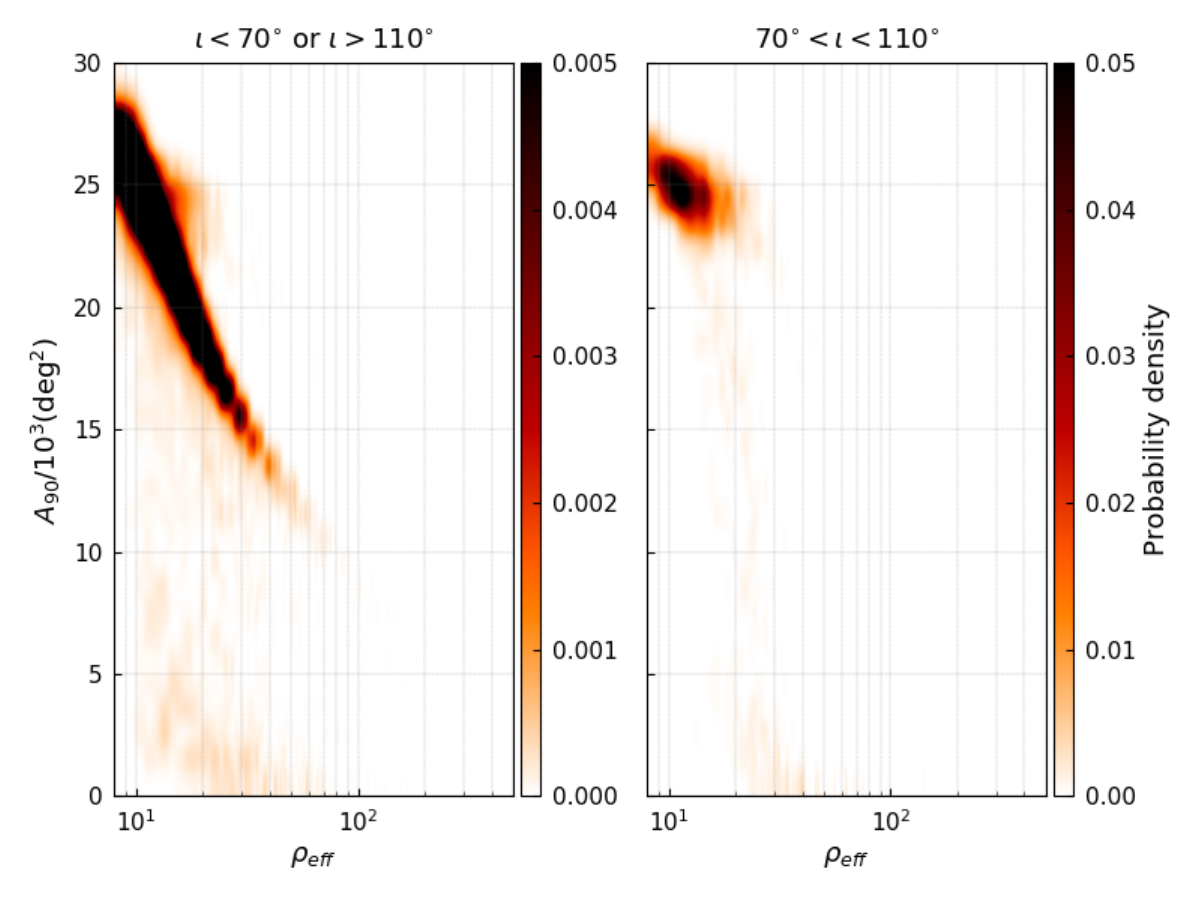}}
\subfloat[\label{fig:area_red_inc}]{\includegraphics[width=\columnwidth]{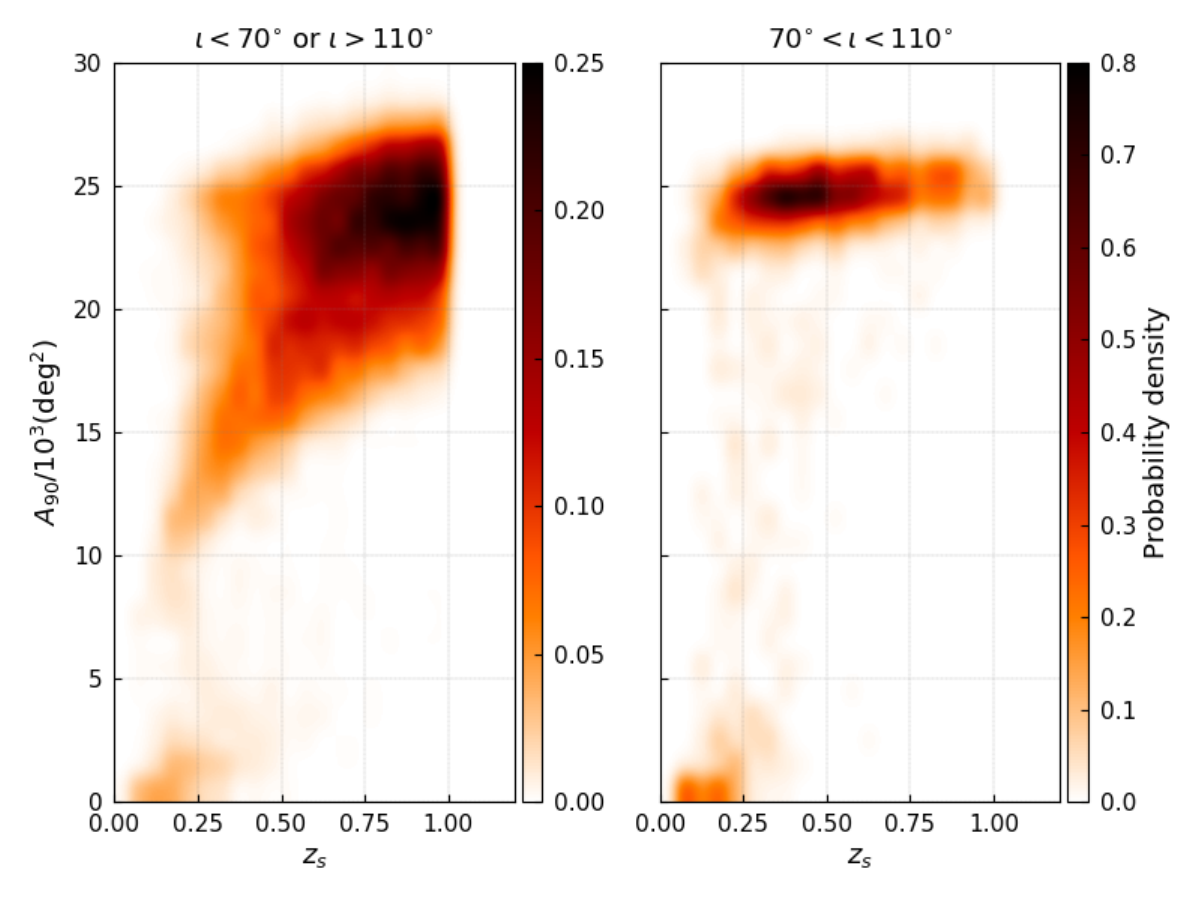}} \\
\subfloat[\label{fig:area_red_chm}]{\includegraphics[width=\columnwidth]{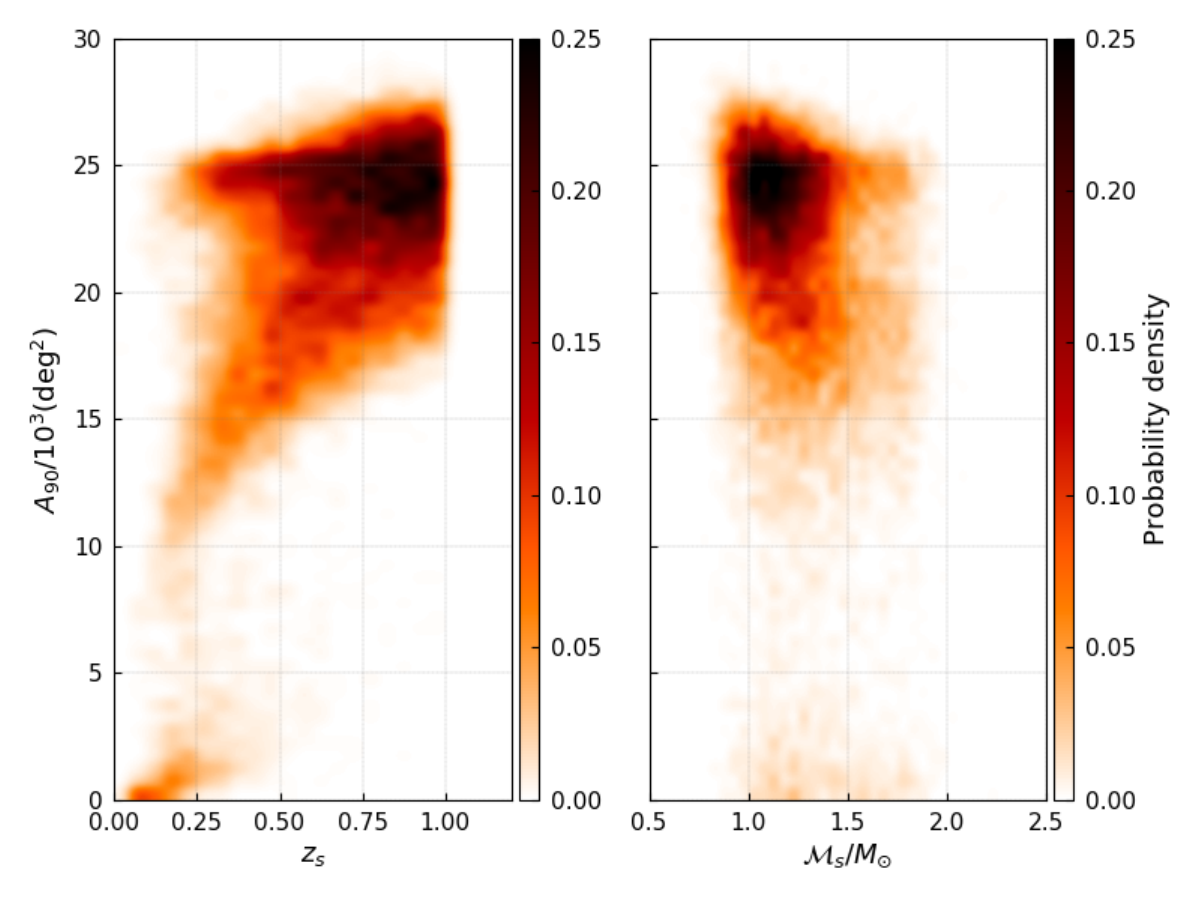}}
\caption{ The dependence of sky localization on $\rho_{eff}$  and redshift is shown in (a) and (b) for two sets of inclination angles, $(\iota < 70^{\circ}$ or $\iota> 110^{\circ})$ and $(70^{\circ}< \iota < 110^{\circ})$. The dependence of sky localization on redshift and chirp mass for all the detected sources is shown in (c).}
\label{fig:loc_rho_inc}
\end{figure*}

The dependence of sky localization on the effective SNR is shown in Fig. \ref{fig:area_rho_inc}. It shows that for the sources which have inclination $\iota < 70^{\circ}$ or $\iota> 110^{\circ}$, 
the area of sky localization decreases exponentially with the effective SNR, while only a small fraction of sources having $ 70^{\circ}< \iota < 110^{\circ}$ 
are detectable. We investigated different cutoffs in the inclination angle and found that given the low SNR values, the sources with $ 70^{\circ}< \iota < 110^{\circ}$ have much poorer localization. No such dependence of sky localization on angle of declination, right ascension, or angle of polarization $\delta, \alpha, \psi$ was seen. Figure \ref{fig:area_red_inc} and \ref{fig:area_red_chm} show that in the mass range of the detected population, irrespective of value of the inclination angle and the chirp mass, only the sources located at $z \lesssim 0.15$ can be localized within a 90\% credible region of 1000 square degrees.

\subsubsection{Mass and distance estimates}\label{mass_z_estimates}

In this analysis, we recover the probability distributions for chirp mass $\mathcal{M}$, redshift $z$, luminosity distance $D_L$, and mass ratio $q$ for 17 994 detected sources in addition to their angular distributions. The relative error on the parameters are estimated from the spread of $90\%$ probability about the median of the recovered distributions for the respective parameters.

\begin{figure*}
\subfloat[\label{fig:err_chm}]{\includegraphics[width=\columnwidth]{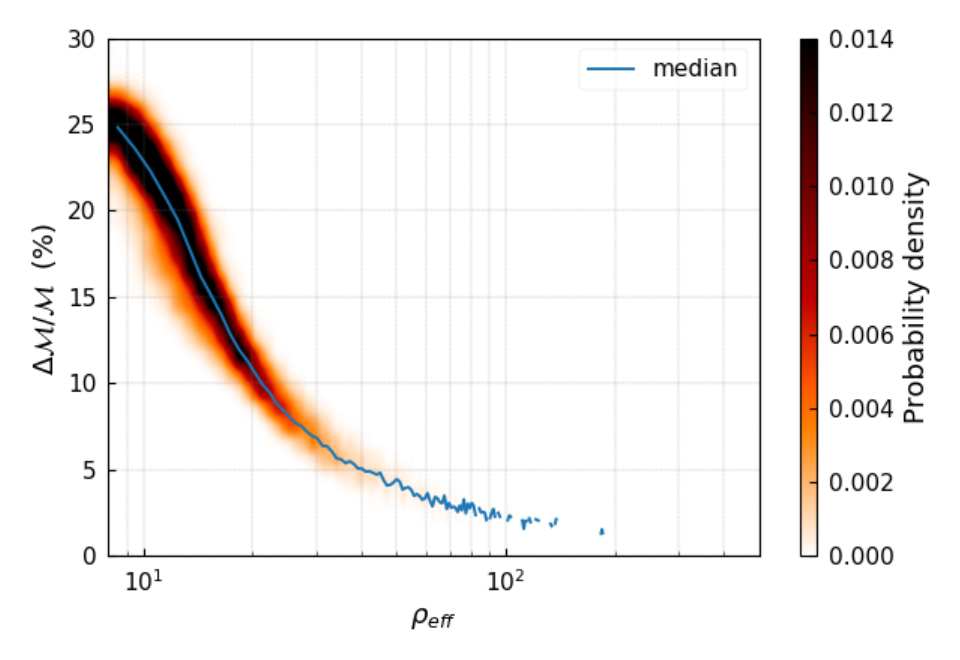}}
\subfloat[\label{fig:err_z}]{\includegraphics[width=\columnwidth]{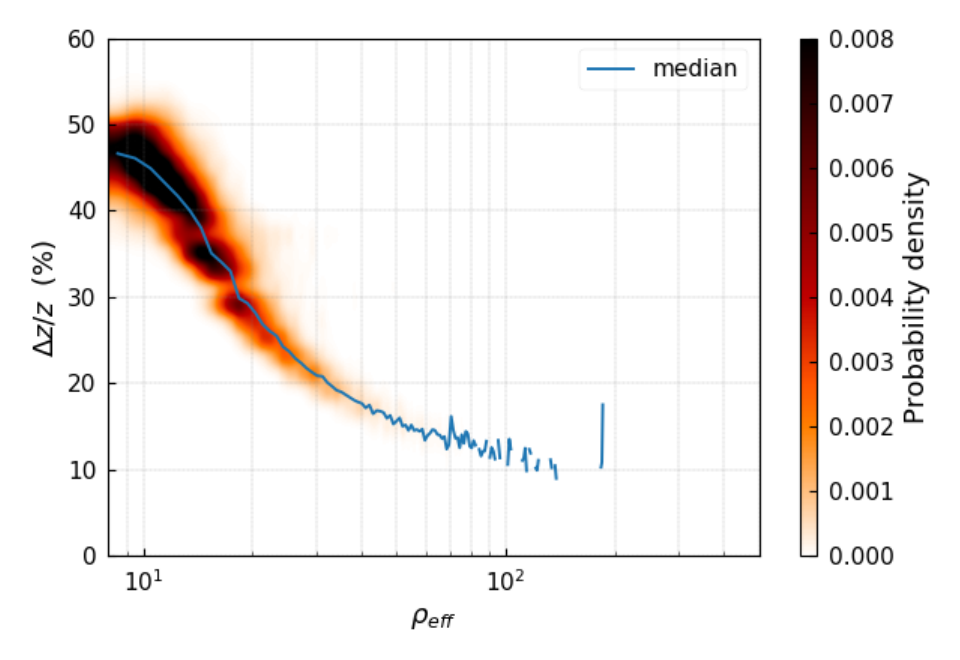}}\\
\subfloat[\label{fig:err_M}]{\includegraphics[width=\columnwidth]{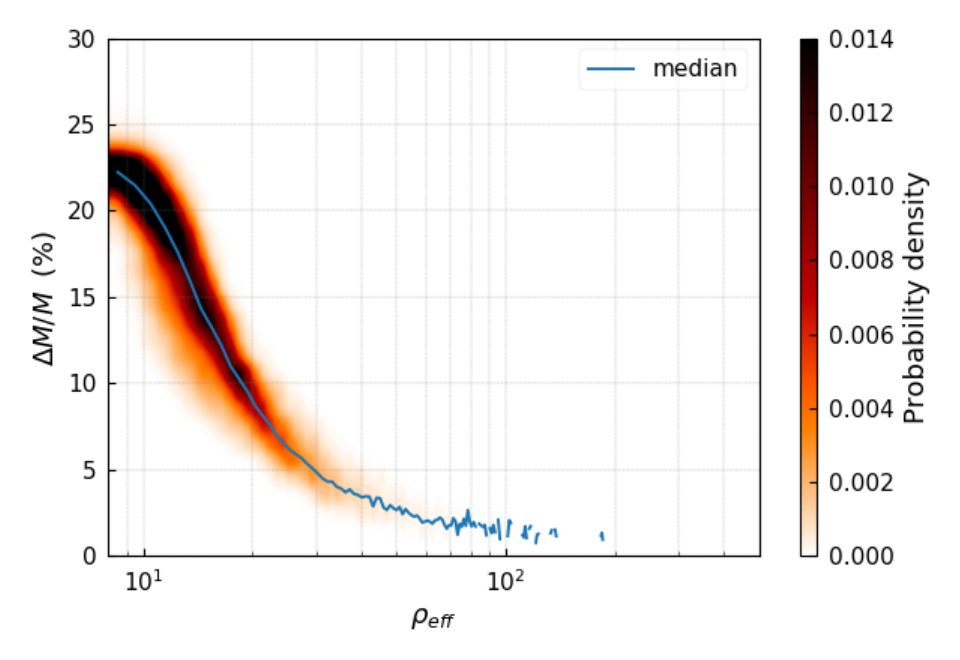}}
\subfloat[\label{fig:err_dl}]{\includegraphics[width=\columnwidth]{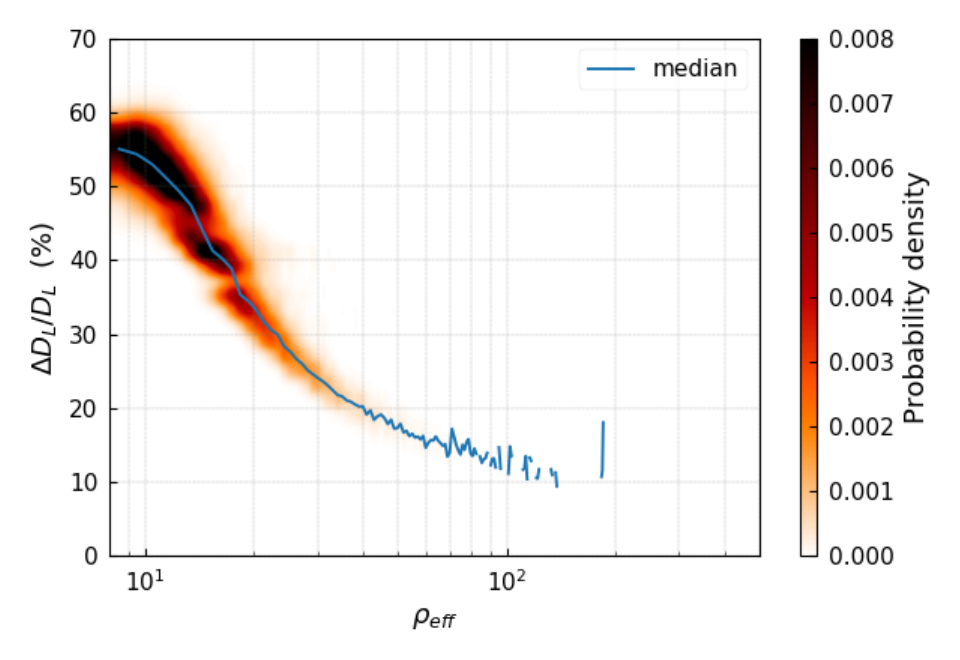}}\\
\subfloat[\label{fig:err_q}]{\includegraphics[width=\columnwidth]{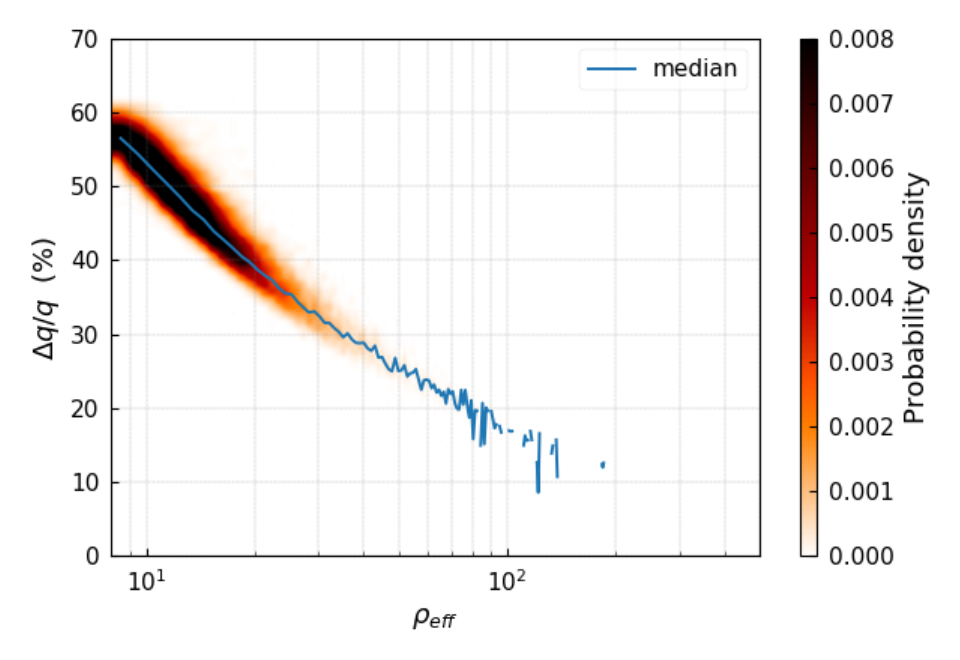}}
\caption{ Distribution of relative errors on (a) chirp mass $\mathcal{M}$, (b) redshift $z$, (c) total mass $M$, (d) luminosity distance $D_L$, and (e) mass ratio $q$, for all the detected sources, with $\rho_{eff}$. The relative error on the parameters are estimated from the spread of $90\%$ probability about the median of the distribution of the respective parameters.}
\label{fig:err_param}
\end{figure*}

Figure \ref{fig:err_param} shows the distribution of relative error on the parameters of the detected compact binary system  with the accumulated effective SNR. We see that the relative error drops with increasing $\rho_{eff}$. For $\mathcal{M}$ it reduces from $\sim 20\%$ at $\rho_{eff} \sim 15$ to $\sim 5\%$ at $\rho_{eff} \sim 50$. Figure \ref{fig:err_z} shows that the relative error redshift goes down from $\sim 40\%$ at $\rho_{eff} \sim 15$ to $\sim 15\%$ at $\rho_{eff} \sim 50$. The respective errors for total mass $M$, luminosity distance $D_L$, and mass ratio $q$ are shown in Figs. \ref{fig:err_M}, \ref{fig:err_dl}, \ref{fig:err_q} respectively.

\begin{figure*}
\subfloat[\label{fig:er_inc_chm}]{\includegraphics[width=\columnwidth]{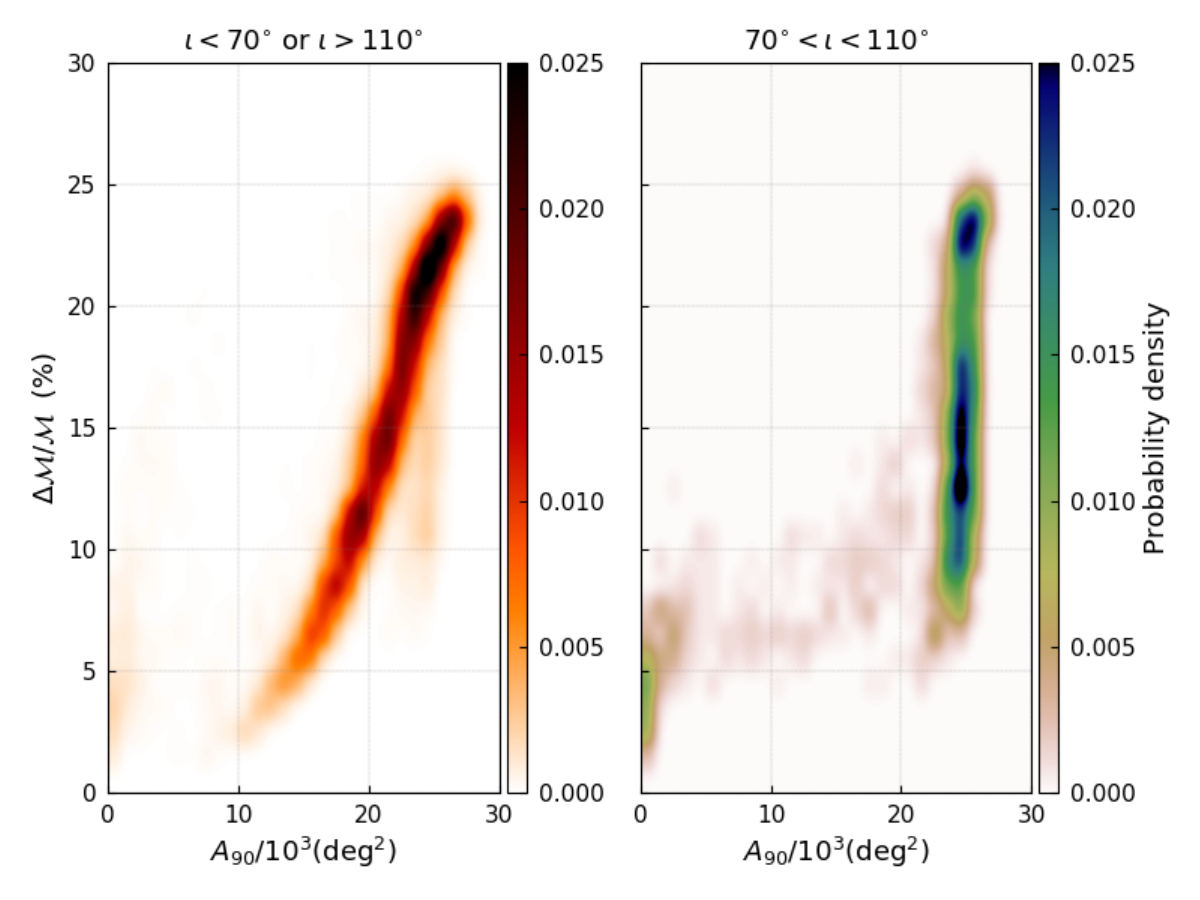}}
\subfloat[\label{fig:err_inc_z}]{\includegraphics[width=\columnwidth]{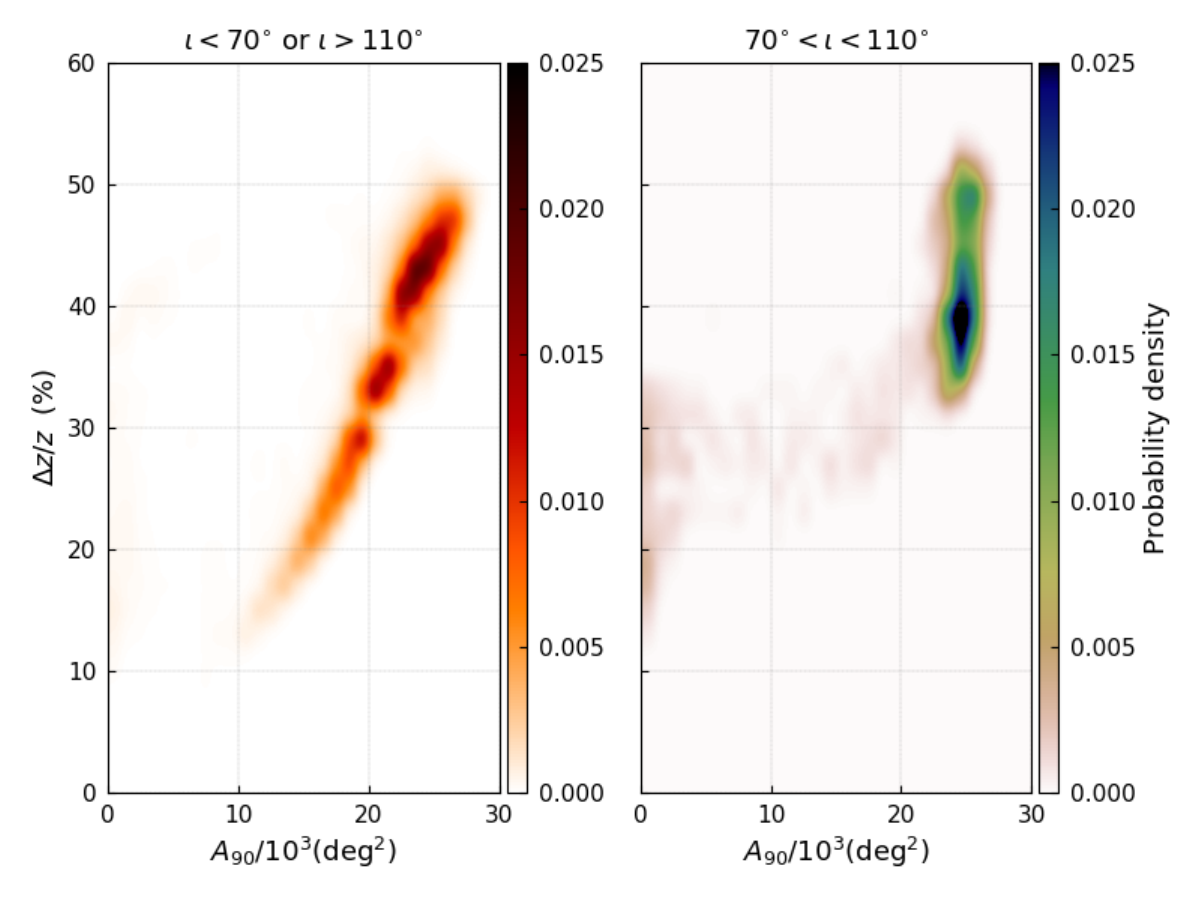}}
\caption{ Distribution of relative errors on (a) chirp mass $\mathcal{M}$, (b) redshift $z$ for all the detected sources with respect to the sky localization for two sets of inclination angles, $(\iota < 70^{\circ}$ or $\iota> 110^{\circ})$ and $(70^{\circ}< \iota < 110^{\circ})$.}
\label{fig:error_inclination}
\end{figure*}

\subsubsection{The effect of inclination on the recovery of parameters}\label{sec:inc_effect}

The discussion in the previous section showed that sky localization has a dependence on the inclination angle $\iota$ of the detected compact binary system. It was shown in Fig. \ref{fig:loc_rho_inc} that, for the detected compact binary sources located at $z \gtrsim 0.15$, the sources with $ 70^{\circ}< \iota < 110^{\circ}$ have a poor localization given the low SNR value. In addition to this, Fig. \ref{fig:error_inclination} shows the distribution of relative errors on $\mathcal{M}$ and $z$ for all the detected sources with the sky localization. The left panels in Figs. \ref{fig:er_inc_chm} and \ref{fig:err_inc_z} show the sources with $\iota < 70^{\circ}$ or $\iota> 110^{\circ}$, and the right panels show the sources with $ 70^{\circ}< \iota < 110^{\circ}$. It can be seen that although only a few sources with $ 70^{\circ}< \iota < 110^{\circ}$ generate enough SNR to be detected, the accuracy of recovered values of $\mathcal{M}$ and $z$ for these low SNR cases, i.e $A_{90}> 23 \times 10^3$ square degrees, is better than for those with $\iota < 70^{\circ}$ or $\iota> 110^{\circ}$. Most of the sources detectable with higher SNR and having better localization are the sources with the inclination angles $\iota < 70^{\circ}$ or $\iota> 110^{\circ}$.

\begin{figure*}
\subfloat[\label{fig:area_combined_cummu}]{\includegraphics[width=\columnwidth]{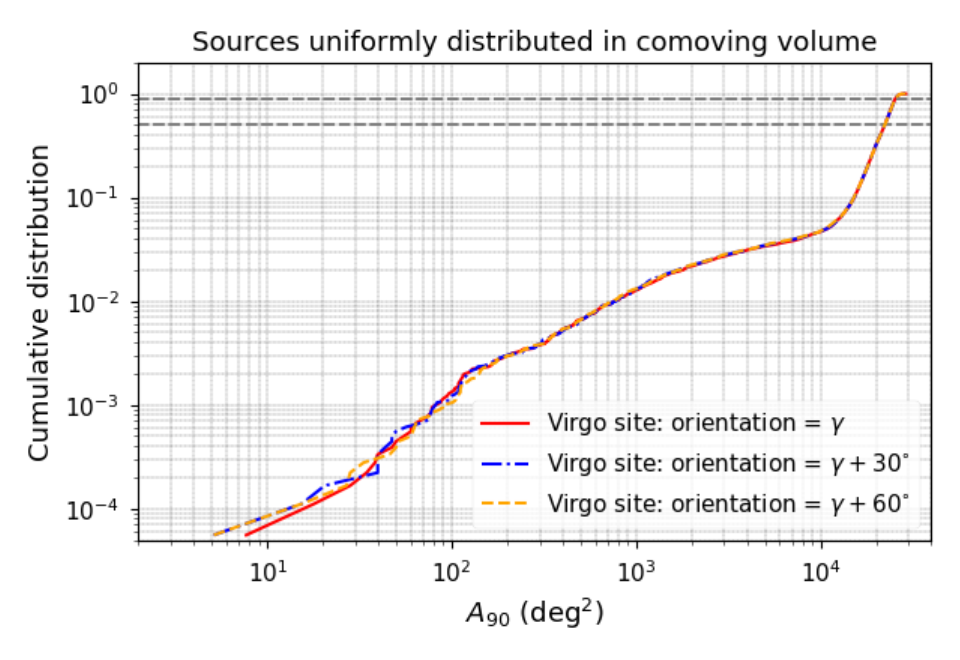}}
\subfloat[\label{fig:area_site}]{\includegraphics[width=\columnwidth]{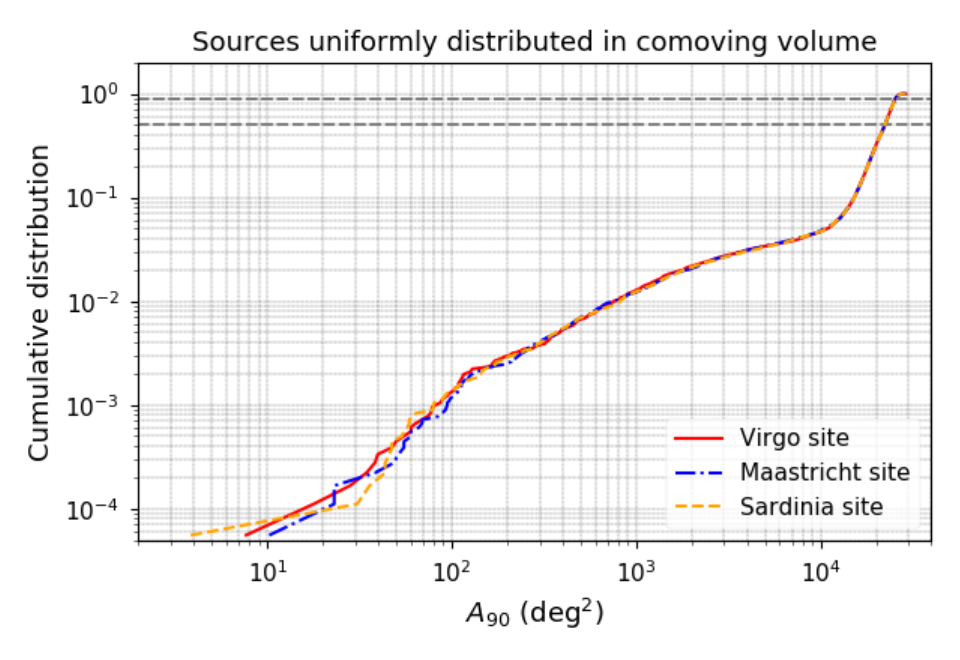}}
\caption{ Plot showing the effect of different site coordinates on the sky localisation. (a) The change in the recovered sky localisation with the change in the orientation of the ET detector located at the Virgo site. Assuming that the initial orientation of ET is $\gamma$, the plot compares the localisation with that obtained for $\gamma +30^{\circ}$ and $\gamma +60^{\circ}$. (b) The recovered sky localisation for Virgo, Maastricht and Sardinia sites. }
\label{fig:loc_site}
\end{figure*}

\subsubsection{The effect of choice of site on the localization capability of ET}

We now proceed to test the localization capability of the ET for different orientations and at different sites. In order to see the affect of orientation of the detector on the sky localization, we assume that the ET is located at the Virgo site and repeat the analysis using the ET detector with three different orientations. The orientation $\gamma$ is measured counterclockwise from East to the bisector of the interferometer arms.  We assume the initial orientation to be $\gamma =  84.84^{\circ}$ and then analyze the same set of detected sources with the ET detector orientated at, $\gamma_1 = \gamma + 30^{\circ}$ and $\gamma_2 = \gamma + 60^{\circ}$. Figure \ref{fig:area_combined_cummu} shows the cumulative distribution of sky localization for these three different cases of orientation of the detector at the Virgo site. It can be concluded that the change in orientation does not have much effect on the sky localization.

We also investigate the change in the sky localization for different site locations of ET. We assume the locations to be in Maastricht and Sardinia in addition to the Virgo site. The coordinates of the sites are mentioned in Table \ref{tab:site_coordinates}. The site locations are chosen keeping in mind that these are the official candidate sites for the ET. The cumulative distributions of sky localization for these locations are shown in Fig. \ref{fig:area_site} and show no change in the effective sky localization.

\begin{figure*}
\subfloat[\label{fig:injchmhist}]{\includegraphics[width=\columnwidth]{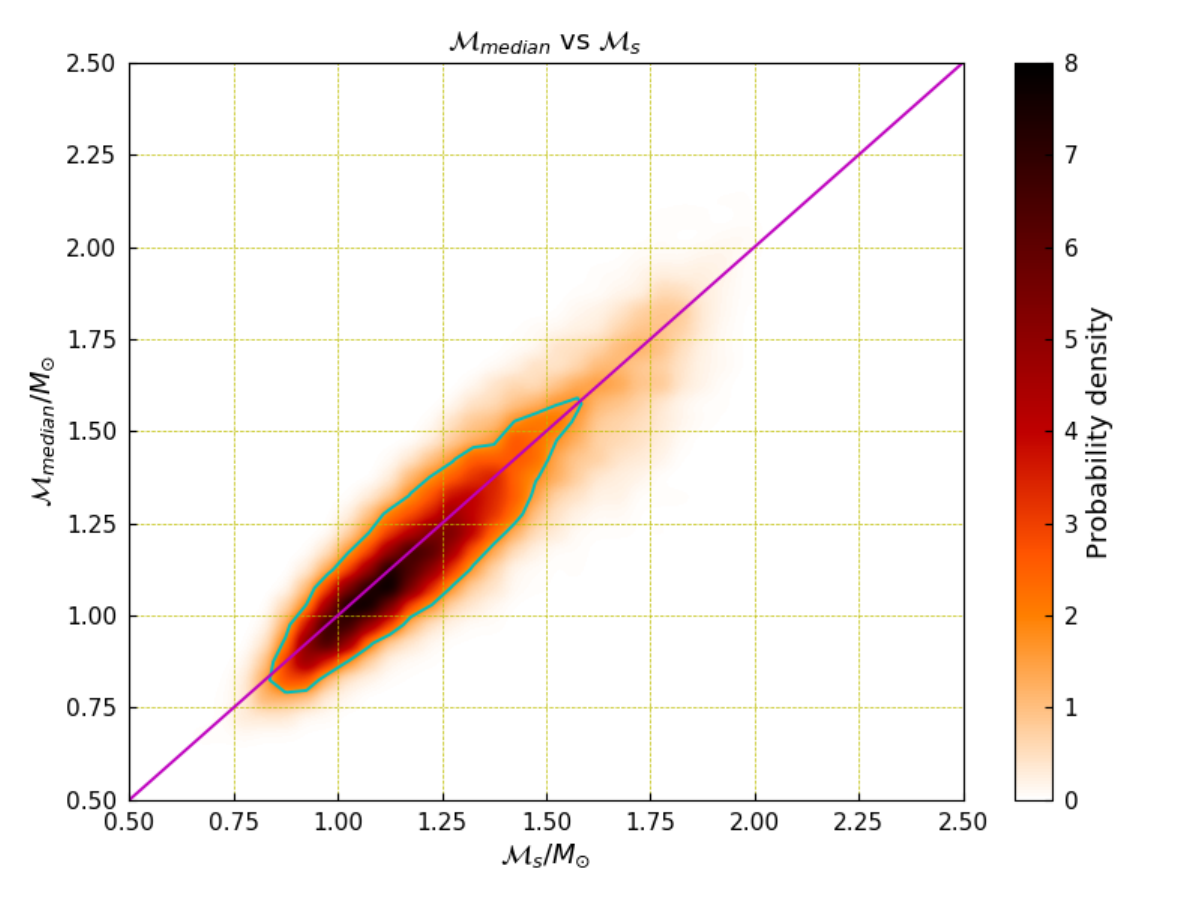}}
\subfloat[\label{fig:injzhist}]{\includegraphics[width=\columnwidth]{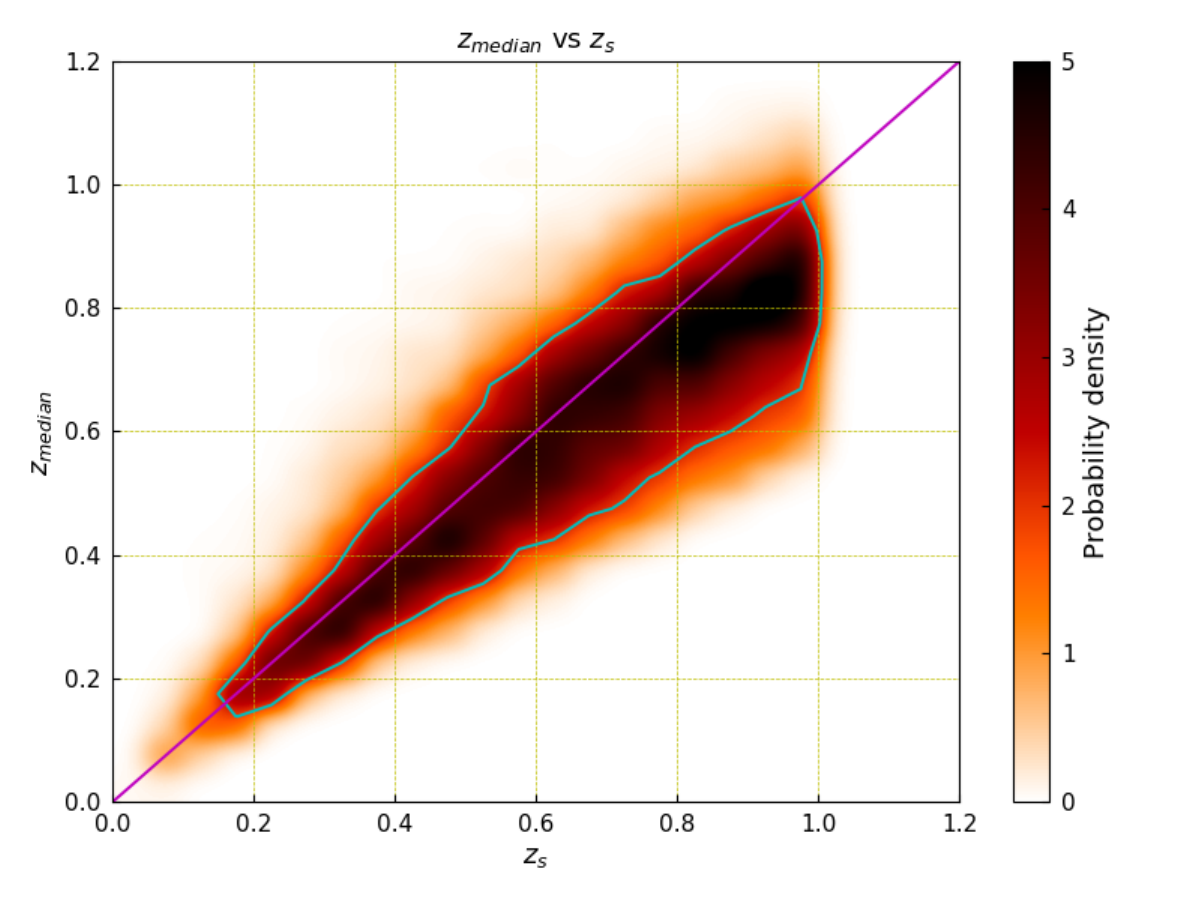}}\\
\subfloat[\label{fig:injMhist}]{\includegraphics[width=\columnwidth]{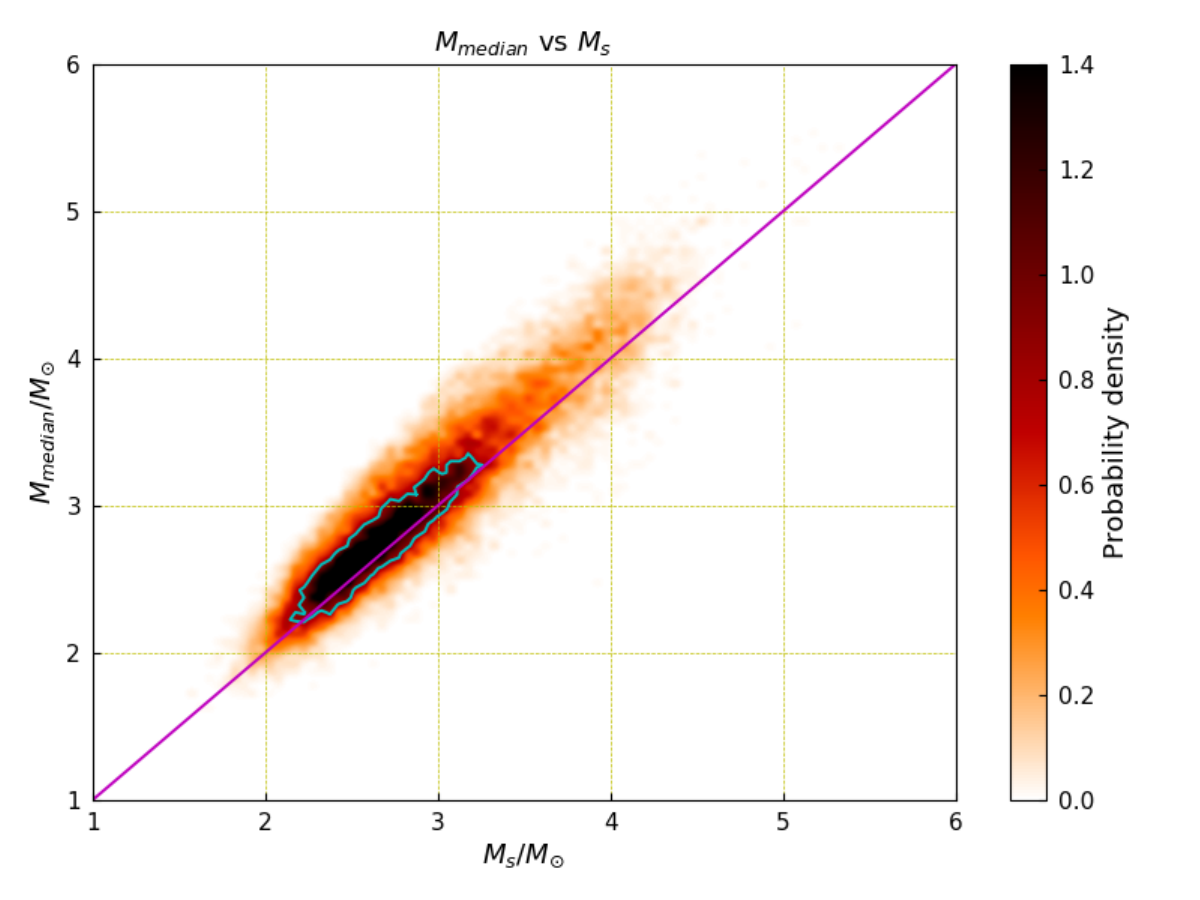}}
\caption{ The density distribution of the recovered median values of (a) $\mathcal{M}$, (b)   $z$, and (c) $M$, w.r.t the actual source values of the respective parameters. The purple lines represents the source$=$median reference line. }
\label{fig:injvsmedian}
\end{figure*}

\subsubsection{Biases in recovery of the parameters of the binary}

In order to check the accuracy of the algorithm in recovering the distributions of the binary parameters, we show the recovered median values of the distributions with respect to source values for chirp mass $\mathcal{M}$, redshift $z$, and total mass $M$ in Fig. \ref{fig:injvsmedian}. The blue contour in the plots encloses the 90\% probability region.

Figure \ref{fig:injchmhist}  shows the distribution of recovered median values of the intrinsic chirp mass $\mathcal{M}_{median}$ with respect to the actual source value $\mathcal{M}_s$ values. It is seen that the chirp masses are recovered correctly over the whole range of source values.

In the case of the median values of the redshift distribution, shown in Fig. \ref{fig:injzhist}, we see that all cases detected with $z \gtrapprox 0.6$ are slightly underestimated. This can be explained due to the low SNR generated by these sources and, hence, the larger error associated with the SNR as per our prior assumption. A more detailed explanation of the origin of this bias is given in the Appendix.

Figure \ref{fig:injMhist} shows the recovered median vs source value for the total mass $M$. We see that the recovered median values are overestimated over the whole range of source values. As mentioned earlier, we assume that $f_{max}$, which gives the value of redshifted total mass, is measured accurately. Since redshift is underestimated, we see the overestimation in the recovered total mass $M$ values.

\section{Conclusion}

In this analysis, we studied the ability of ET as a single instrument to study longer-duration signals from coalescing low mass compact binary systems. We assume the detector to be located at the Virgo site and analyze the signal every 5 min, assuming that the response functions for the three ET detectors do not change much within that period.

We show that, although one cannot use time of flight delays to constrain the position in the sky of a given source in case of three co-located detectors in single ET, combining information from different antenna patterns for each of the three detectors in each time segment provides good constraints on the parameters of a merging binary system.

We analyzed a mock population of compact binary sources for which the signal will stay for a long duration in the ET detection band. Assuming that the change in the response functions is negligible within 5 min, we divided the inspiral signal into 5 min segments from the time it enters the detection band of ET at 1Hz with the duration of the last segment limited by $f_{max}$, the frequency at the end of the inspiral. 

We assumed the threshold of detection to be $\rho^i_j > 3$ for the $i^{th}$ segment in the $j^{th}$ detector in at least one segment, for $j = (1,2,3)$ corresponding to the three ET detectors comprising single ET and the accumulated effective SNR $\rho_{eff} > 8 $.

The angles describing the location of the source, its inclination and polarization are constrained using the ratios of the SNRs generated in each signal segment of the three detectors of the equilateral triangle configuration of ET. This, in turn, provide constraints on the antenna response function $\Theta^i_{eff}$ in the $i^{th}$ segment.

We then use the information about $\Theta^i_{eff}$, $\rho^i_{eff}$ and initial and final GW frequency $f_{i-1}, f_i$ in $i^{th}$ segment to estimate the parameter $\Lambda$. Combining the information about the angles from each segment increases the accuracy of the estimate of the angles and also gives a stricter constraint on $\Lambda$. We then use the constraint on $\Lambda$ to estimate intrinsic chirp mass $\mathcal{M}$, redshift $z$, total mass $M$, luminosity distance $D_L$, and mass ratio $q$ of the merging binary system. 

We conclude that the ET as a single instrument can localize the low mass compact binary sources and break the chirp mass - redshift degeneracy.  The analysis presented here allows us to estimate source frame masses and redshifts of the coalescing compact binaries which facilitates the population study of compact object binaries \cite{2022A&A...667A...2S}.

We find that the accuracy of determination of the redshift and the source frame chirp mass with ET as a single instrument, is typically 40\% and 20\% , respectively for $\rho_{eff} \sim 15$ and it is $\sim 15\%$ and $\sim 5\%$ for $\rho_{eff} \sim 50$. In the best case we see that using this method for analyzing a long-duration signal, single ET in triangular configuration can constrain the localization area for 90\% probability region of $(\delta, \alpha)$ to a minimum value of 7.68 square degrees, for $\rho_{eff} = 184.58$, although only $\approx 1\%$ of binaries can be localized with 90\% credibility, within 800 square degrees. It should be noted that the dominant error in the analysis is the one on the SNRs and our assumption of $\sigma_{\rho} = 1$ is a conservative one. 

We also studied the effect of orientation of the detector on the sky localization with three different orientations of ET detector assumed to be located at the Virgo site and found that the change in orientation has no effect on the sky localization. In addition to this we also investigated the change in the sky localization for different site locations of ET. We assumed the location of ET to be in Maastricht and Sardinia in addition to Virgo, and found no change in the effective sky localization due to change in the ET site for these sites.

\begin{acknowledgments}
We acknowledge the support from the Foundation for Polish Science Grant No. TEAM/2016-3/19 and NCN Grant No. UMO-2017/26/M/ST9/00978. N.S. is supported by the "Agence Nationale de la Recherche”, Grant No. ANR-19-CE31-0005-01 (PI: F. Calore). We thank the anonymous reviewer for all the valuable comments and suggestions which helped us to improve the manuscript. This document has been assigned Virgo document number VIR-0785A-21.

\end{acknowledgments}

\appendix

\section{Appendixes}\label{sec:append}

\begin{figure*}
\centering
\includegraphics[width = 1.5\columnwidth]{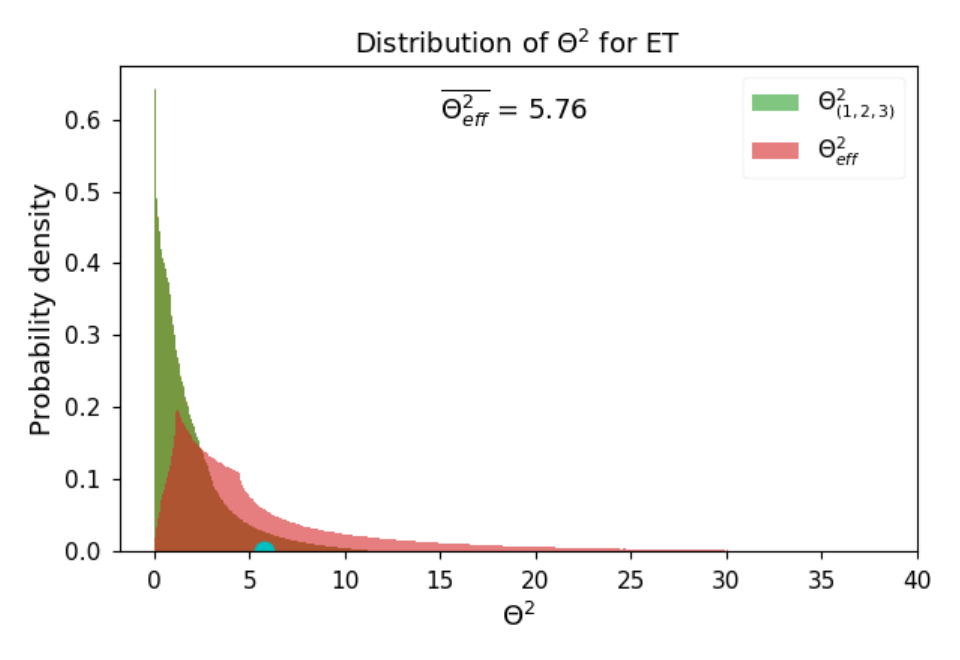}
\caption{The plot shows the probability distributions for $\Theta^2$ for each of the three ET detectors and the probability distribution of $\Theta^2_{eff}$. The blue dot denotes the mean value $\overline{\Theta^2_{eff}}$.}\label{fig:theta_dist_ET}
\end{figure*}

In Fig. \ref{fig:injchmhist}, we see the deviation of the median value of the parameter from the true value of that parameter. This origin of this \emph{bias} can be understood from the Fig. \ref{fig:theta_dist_ET} which shows the distribution of $\Theta$ defined in Eq. (\ref{theta}), assuming that the probability distributions of angular distribution $\cos\delta, \alpha/ \pi$, $\cos \iota$ and $\psi/ \pi$, are uncorrelated and are distributed uniformly over the range $[-1,1]$. We see that this distribution of $\Theta$ is inherently biased toward lower values. Therefore, for a given $\rho$, defined in Eq. (\ref{snr}), the smaller $\Theta$ has to be compensated by larger chirp mass $\mathcal{M}$ and smaller redshift $z$. Thus there is an inherent bias toward lower redshift and larger chirp mass.

\begin{figure*}
\centering
\includegraphics[width = 1.5\columnwidth]{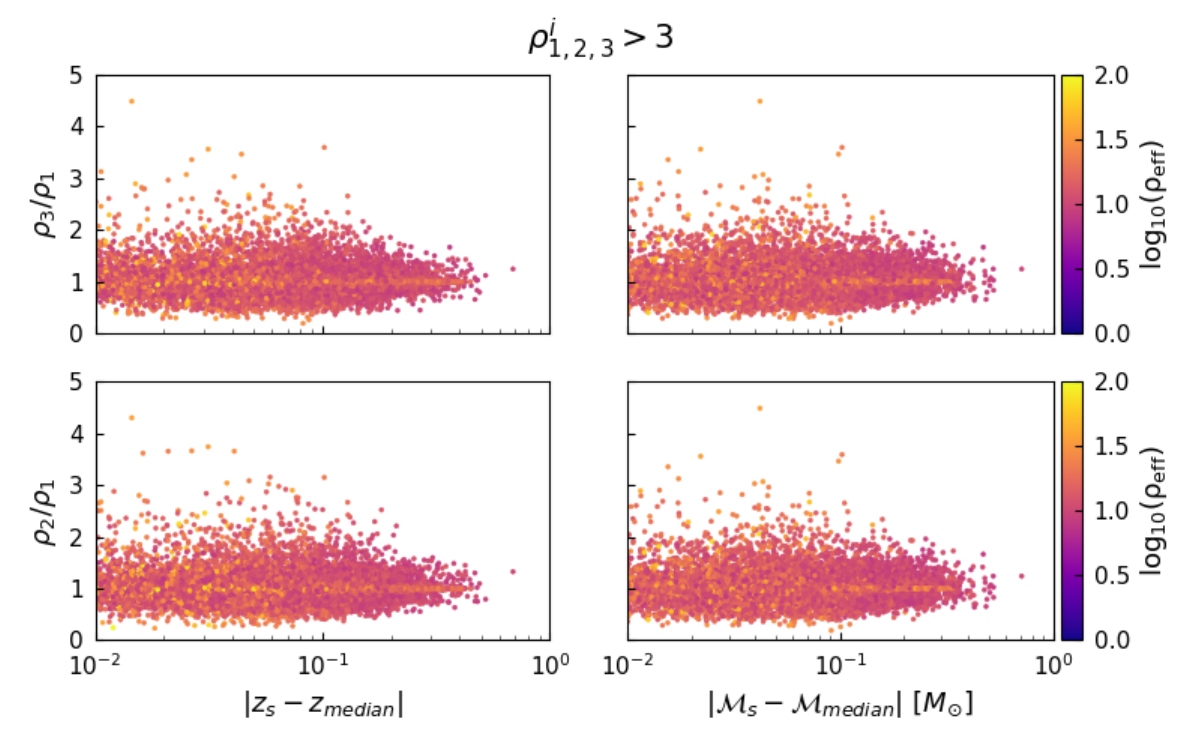}
\caption{The left panel shows the absolute value of bias in redshift as a function of ratios of SNRs, and the right panel shows the absolute value of bias in chirp mass as a function of ratios of SNRs.}
\label{fig:ratio_bias}
\end{figure*}

One of the main building blocks of our analysis is that we use the ratios of SNRs in each segment for each of the three detectors in the triangular configuration of ET, to constrain the value of the effective antenna pattern. This is expressed in Eq. (\ref{theta-ratios}), assuming that the measurement error on the SNRs is Gaussian with the standard deviations for $\rho^i_j$ being $\sigma_{\rho}=1$. In Fig. \ref{fig:ratio_bias}, we show the absolute value of bias as the function of the ratio of the SNRs. We see that the estimate of parameters is likely to be biased for $\rho_2/\rho_1 \approx 1$ and $\rho_3/\rho_1 \approx 1$. The reason for this can be understood by Fig. \ref{fig:theta_dist_ET}. We see that the distributions for $\Theta_1$, $\Theta_2$ and $\Theta_3$ are the same. So in the case of $\rho_2/\rho_1 \approx 1$ and $\rho_3/\rho_1 \approx 1$, there is a much larger number of lower values of $\Theta$ in the probability distribution constrained by these ratios of the SNRs. This results in smaller redshift and higher chirp mass in the respective distributions recovered for these parameters as explained above.

In order to see the variation in this bias due to the inclusion of segments of lower SNRs, we repeated the analysis for the following different detection criteria for 5-min segments.

\begin{itemize}
    \item Criterion 1: A threshold value of accumulated effective SNR $\rho_{eff}>8$ and the SNR for the $i^{th}$ segment in the $j^{th}$ detector {\color{blue}$\rho^i_j> 3$} in at least one segment. This is the criteria which we have used in our main analysis in this paper.
    \item Criterion 2: A threshold value of accumulated effective SNR $\rho_{eff}>8$ and the SNR for the $i^{th}$ segment in the $j^{th}$ detector {\color{blue}$\rho^i_j> 2$} in at least one segment.
    \item Criterion 3: A threshold value of accumulated effective SNR $\rho_{eff}>8$ and the SNR for the $i^{th}$ segment in the $j^{th}$ detector {\color{blue}$\rho^i_j> 1$} in at least one segment.
\end{itemize}

\begin{figure*}
\subfloat[\label{fig:chm_snr_3}]{\includegraphics[width=\columnwidth]{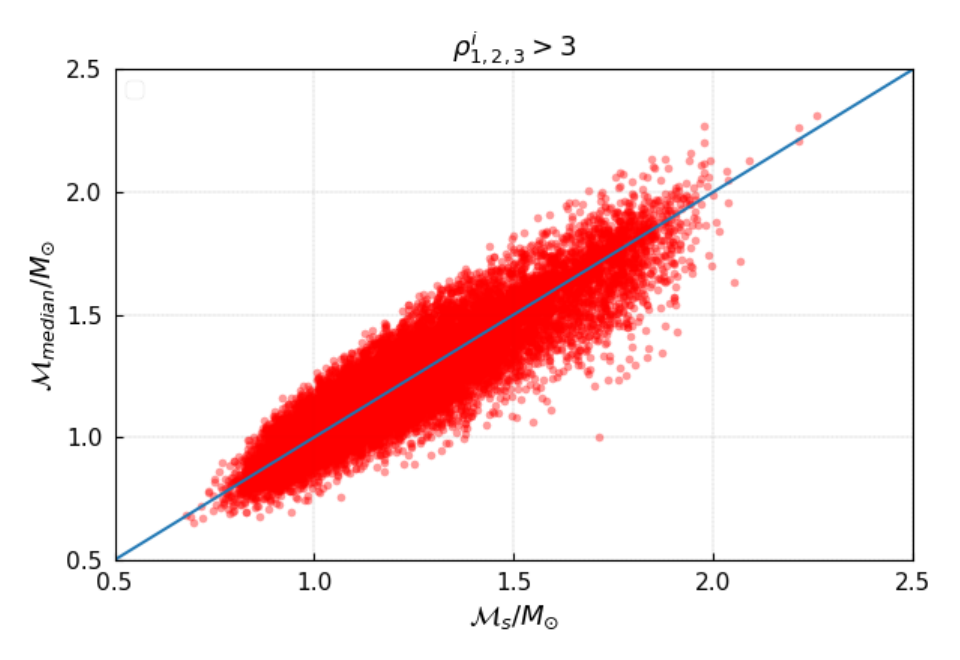}}
\subfloat[\label{fig:z_snr_3}]{\includegraphics[width=\columnwidth]{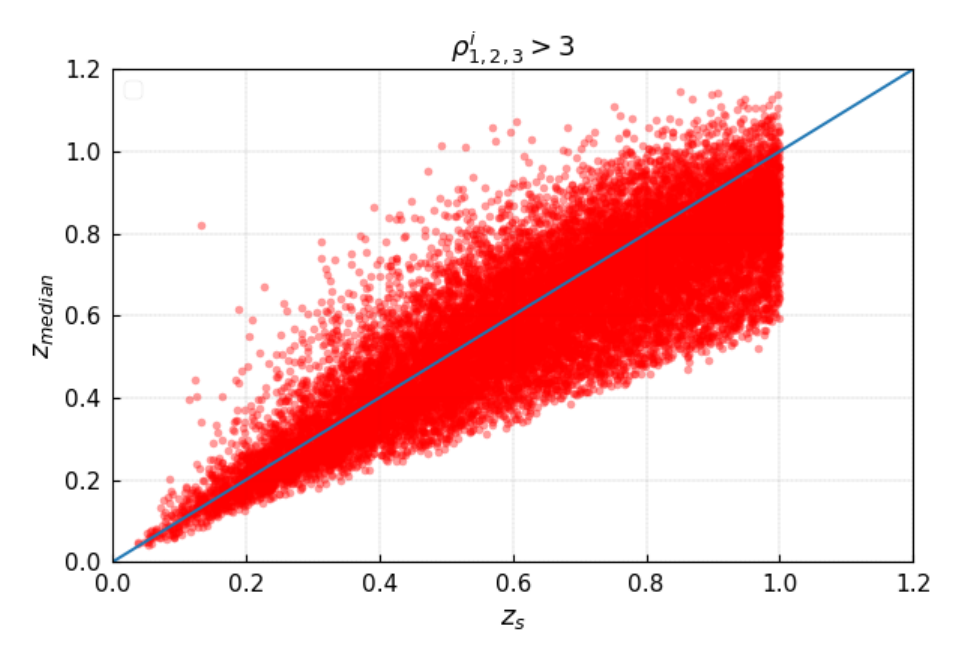}}\\
\subfloat[\label{fig:chm_snr_2}]{\includegraphics[width=\columnwidth]{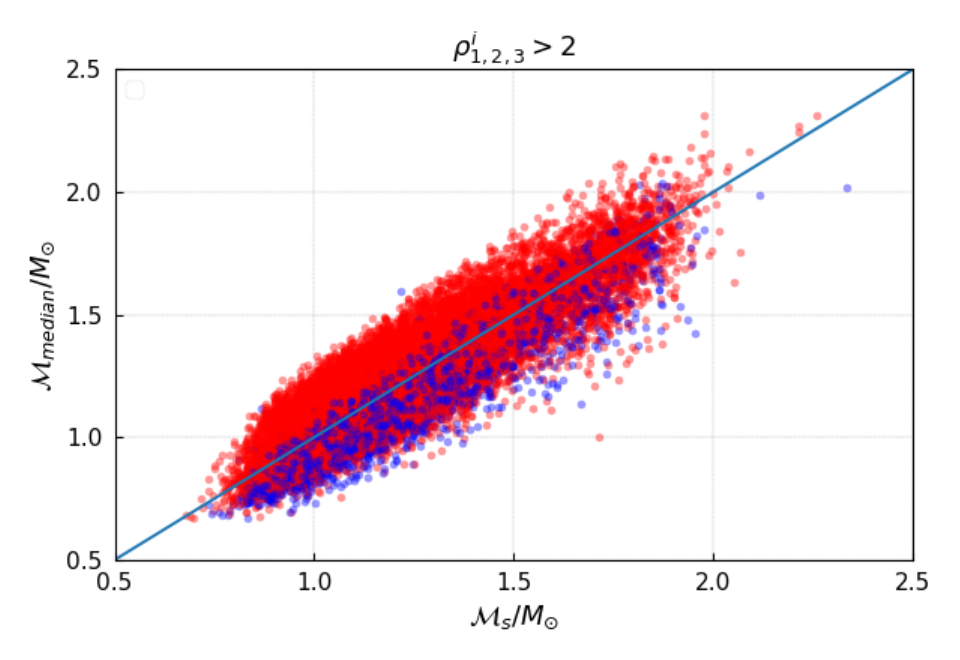}}
\subfloat[\label{fig:z_snr_2}]{\includegraphics[width=\columnwidth]{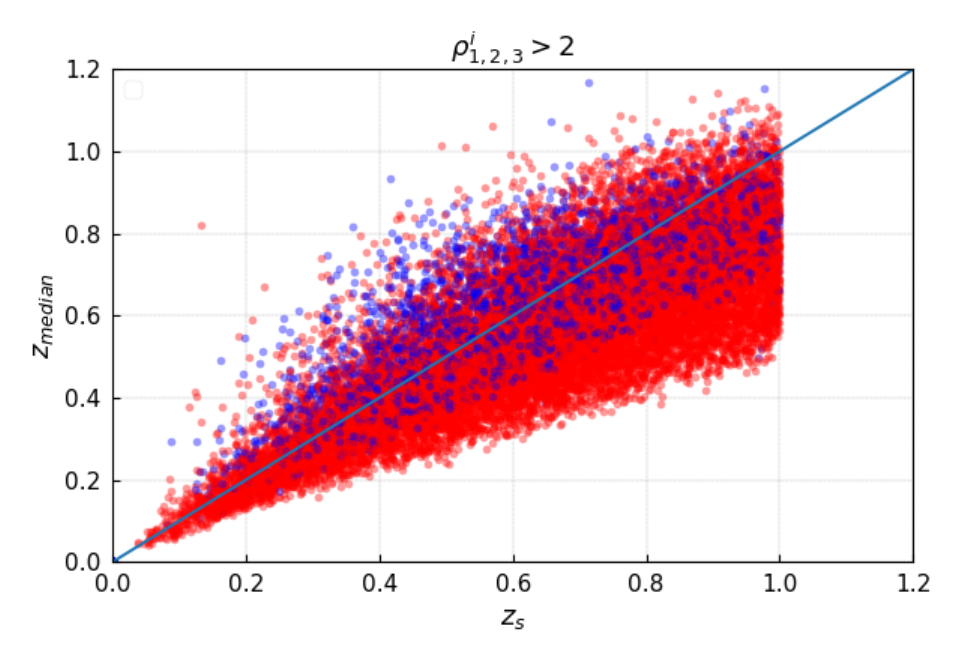}}\\
\subfloat[\label{fig:chm_snr_1}]{\includegraphics[width=\columnwidth]{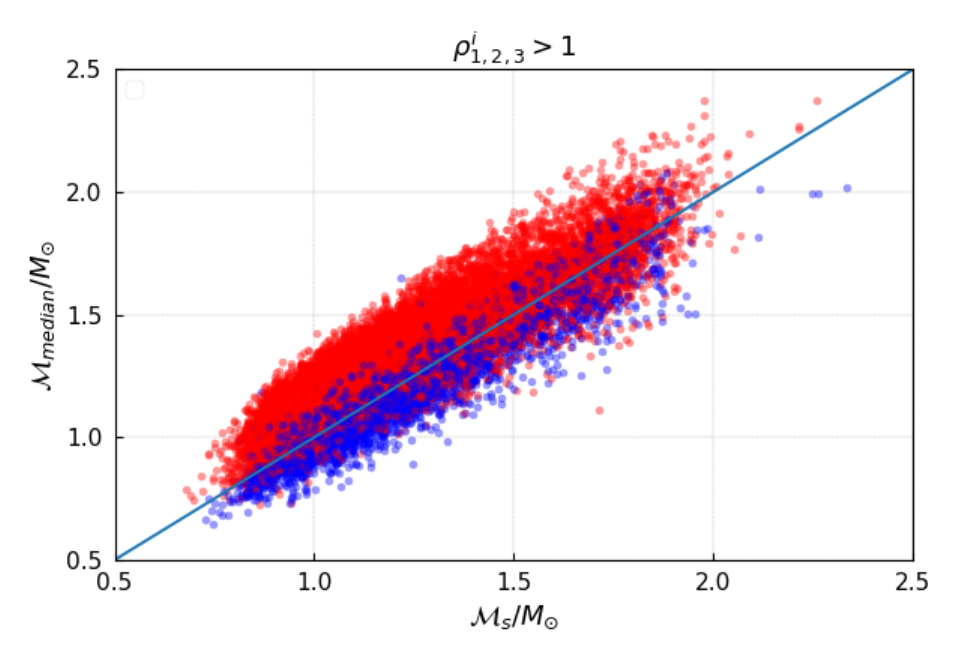}}
\subfloat[\label{fig:z_snr_1}]{\includegraphics[width=\columnwidth]{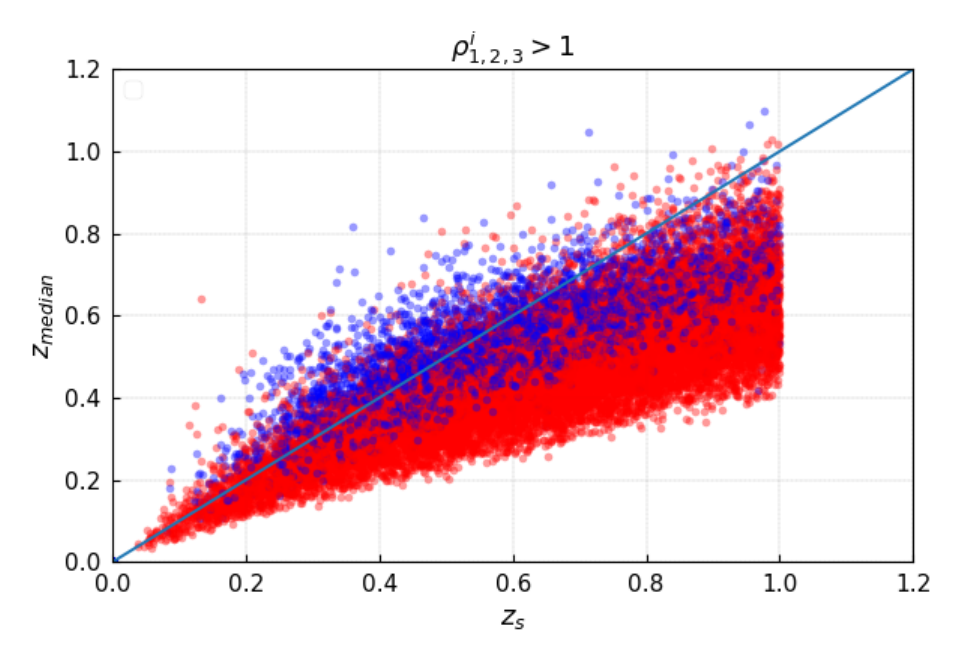}}\\

\caption{The recovered median values vs the actual source values of the chirp mass $\mathcal{M}$ (a,c,d) and redshift $z$ (b,d,f). Top row: sources which are detectable using criterion 1. Middle row: The red point are the ones which were detectable with criterion 1 but were again detected and analyzed using criterion 2. The blue points are the one which were not detectable using criterion 1 but are detectable using criterion 2. Bottom row: The red point are the ones which were detectable with criteria 1 but were again detected and analyzed using criterion 3. The blue points are the one which were not detectable using criterion 1 but are detectable using criterion 3.}
\label{fig:injvsmedian_appendix}
\end{figure*}

The recovered median values vs the actual source values of chirp mass and redshift, using these three detection criteria, are shown in Fig. \ref{fig:injvsmedian_appendix}. The left panel shows the comparison for the chirp mass, and the right panel shows the comparison for the redshift. The top panel shows the comparison of the estimated median value with respect to the actual value for criterion 1. When we repeat the analysis while including segments with $\rho^i_j> 2$, shown in the middle panel, we see that a few more sources cross the detection threshold and are shown in blue. The red points are the sources which were detectable with criterion 1, but their analysis now includes the additional low SNR segments. The bottom row shows the subsequent inclusion of segments of $\rho^i_j> 1$, thus detecting a few more sources, shown in blue. While lowering the detection threshold detects a few more sources, there is also a slight worsening of the bias as we include segments of lower SNRs in the analysis. This can be clearly seen from the red points as we go from the top panel to the bottom panel. The underestimation of redshift and the overestimate of chirp, are both slightly worsened. 

\begin{figure*}
\subfloat[\label{fig:chm_snr2}]{\includegraphics[width=\columnwidth]{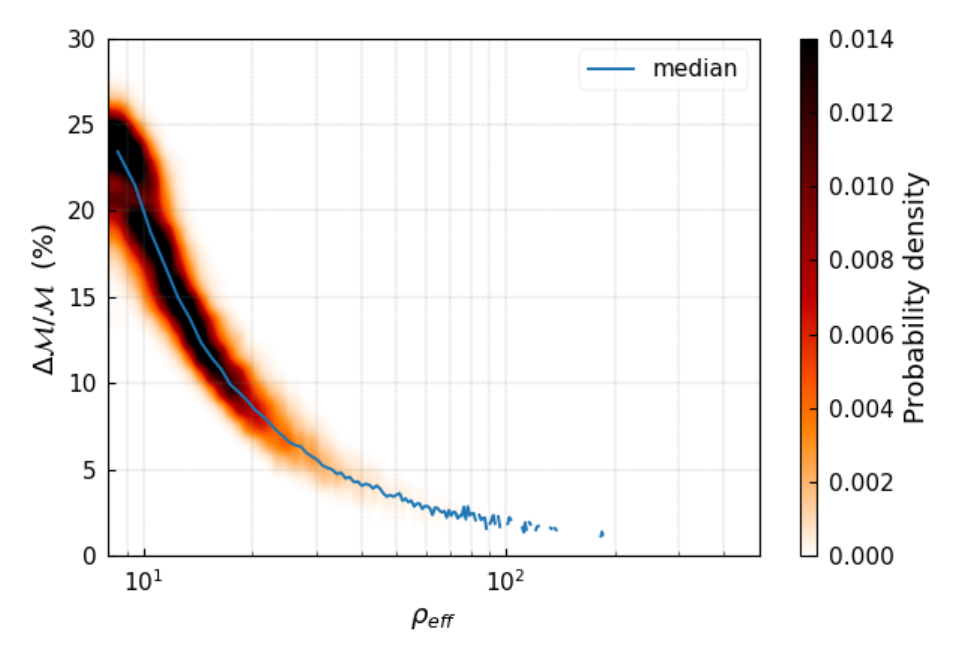}}
\subfloat[\label{fig:z_snr2}]{\includegraphics[width=\columnwidth]{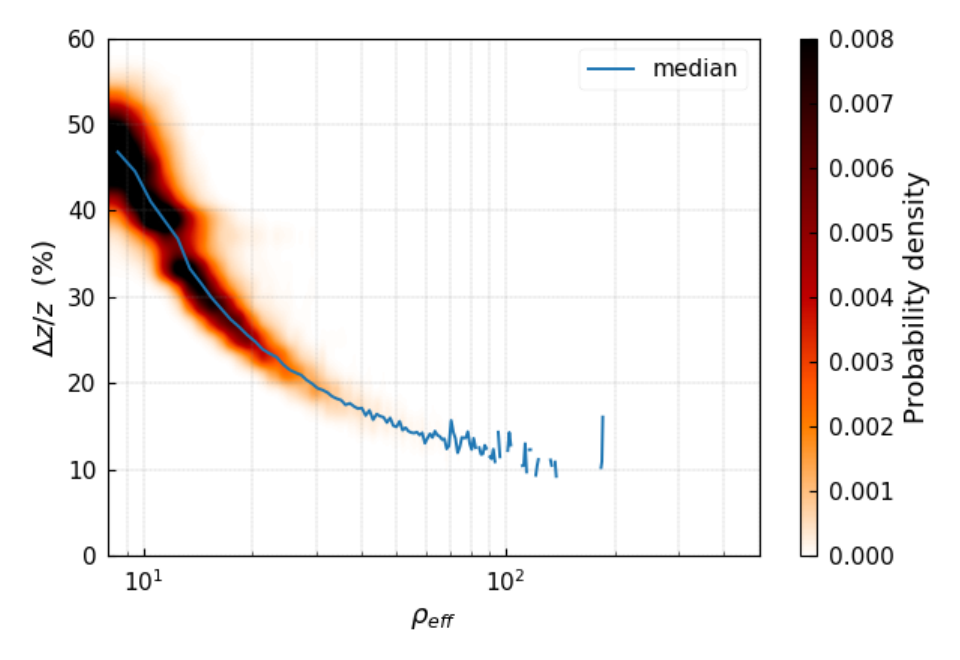}}\\
\subfloat[\label{fig:chm_snr1}]{\includegraphics[width=\columnwidth]{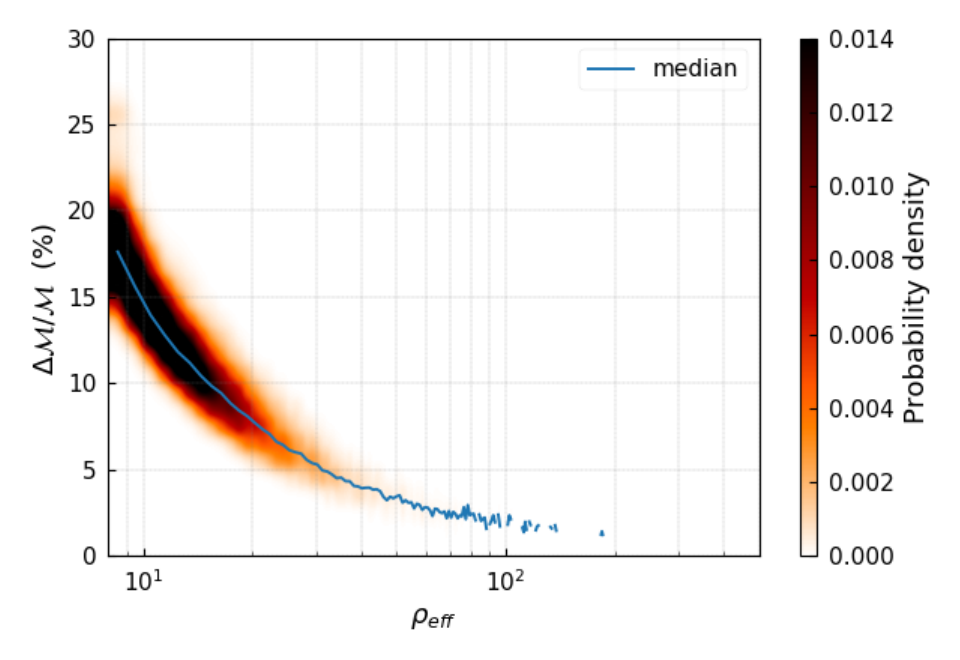}}
\subfloat[\label{fig:z_snr1}]{\includegraphics[width=\columnwidth]{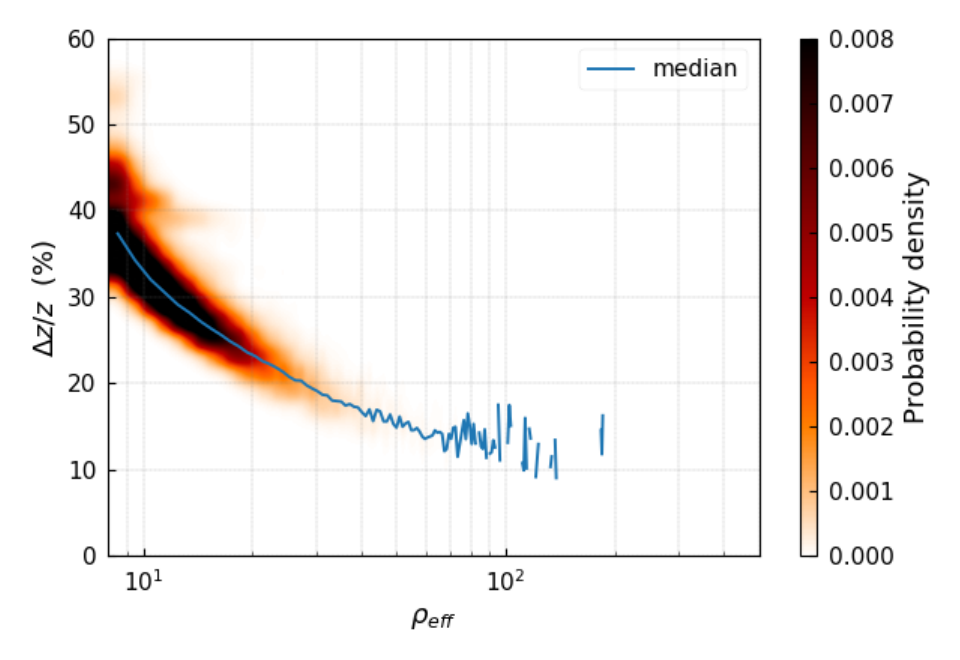}}\\

\caption{Distribution of relative errors on chirp mass $\mathcal{M}$ (left) and redshift $z$ (right), estimated using criterion 2 (a,b) and criterion 3 (c,d).}
\label{fig:relerror_appendix}
\end{figure*}

Although the inclusion of segments of lower SNR worsens the bias, the error on the estimated parameters is lowered due to the additional information from these segments. The distribution of relative errors on chirp mass and redshift estimated using criterion 2 and criterion 3 are shown in Figure \ref{fig:relerror_appendix}. Comparing these with the error estimates obtained using Criterion 1 shown in Fig. \ref{fig:err_param}, we see that the additional segments contribute in reducing the relative errors, mainly for sources generating lower SNRs.

\bibliography{ref}
\bibliographystyle{apsrev}

\end{document}

%% file: macros.tex
%

\newcommand\note[1]{\textcolor{red}{\bf {#1}}}

\renewcommand{\today}{\number\day\space\ifcase\month\or
  January\or February\or March\or April\or May\or June\or
  July\or August\or September\or October\or November\or December\fi
  \space\number\year}
\newcommand\checkme[1]{\textcolor{blue}{#1}}

\def\etc{{\it etc.}}
\def\eg{{\it e.g.}}
\def\vs{{\it vs.}}
\def\etal{{\it et al.}}
\def\ie{{\it i.e.}}
\def\cf{{\it cf.}}
\def\mycaps#1{{\textsc{\small #1}}}
\def\be{\begin{equation}}
\def\ee{\end{equation}}
\def\bi{\begin{itemize}} 
\def\ei{\end{itemize}}
\def\ben{\begin{enumerate}}
\def\een{\end{enumerate}}

\def\aj{Astron. J.}
\def\apj{Astrophys. J.}
\def\apjl{Astrophys. J. Lett.}
\def\apjs{Astrophys. J. Supp. Ser. }
\def\aa{Astron. Astrophys. }
\def\aap{Astron. Astrophys. }
\def\araa{Ann.\ Rev. Astron. Astroph. }
\def\aapr{Astron. Astrophys. Rev. }
\def\physrep{Phys. Rep. }
\def\mnras{Mon. Not. Roy. Astron. Soc. }
\def\mmsun{M_\odot}
\def\prl{Phys. Rev. Lett.}
\def\prd{Phys. Rev. D.}
\def\azh{Soviet Astron.}
\def\apss{Astrophys. Space Sci.}
\def\cqg{Class. Quantum Grav.}
\def\memsai{Mem. Societa Astronomica Italiana}
\def\jcap{Journal of Cosmology and Astroparticle Physics}
\def\nat{Nature}
\def\bain{Bull. Astron. Inst. Netherlands}
\def\planss{Planet. Space Sci.,}
\def\ao{Appl. Opt.,}